\documentclass[amsmath,amssymb,onecolumn,superscriptaddress]{revtex4}

\usepackage{graphicx}%
\usepackage{dcolumn}%
\def\be{\begin{equation}}
\def\ee{\end{equation}}
\def\beq{\begin{eqnarray}}
\def\eeq{\end{eqnarray}}

\usepackage{bm}
\usepackage{graphicx}
\usepackage{dcolumn}
\usepackage{amsmath}
\usepackage[latin1]{inputenc}
\usepackage{graphicx, psfrag}
\usepackage{amssymb}
\usepackage[colorlinks=true, citecolor=blue, urlcolor = blue, linkcolor= red, bookmarks=true]{hyperref}
\usepackage{float}
\usepackage{mathtools}
\usepackage{amsmath}
\usepackage{amsfonts}
\usepackage{dcolumn}
\usepackage{hyperref}
\usepackage{subfigure}
\usepackage{pgfplots}
\usepackage{epstopdf}
\usepackage{booktabs}
\usepackage{orcidlink}
\usepackage{xcolor}

\begin{document}
\title{Implications of Adler-Finch-Skea solution on charged dark energy star satisfying Karmarkar Condition}

\author{Pramit Rej \orcidlink{0000-0001-5359-0655} \footnote{Corresponding author}}
\email[Email:]{pramitrej@gmail.com, pramitr@sccollegednk.ac.in, pramit.rej@associates.iucaa.in}
 \affiliation{Department of Mathematics, Sarat Centenary College, Dhaniakhali, Hooghly, West Bengal 712 302, India}

\begin{abstract}\noindent
A possible approach for preventing compact astrophysical objects from gravitational collapse into singularities is the idea of dark energy. Since it is the cause of our universe's accelerated expansion, it has the greatest impact on the cosmos. As a result, it appears that dark energy can interact with any compact astrophysical stellar object [Phys. Rev. D 103, 084042 (2021)]. In this study, our primary objective is to develop a simpler model of a charged strange star coupled with anisotropic dark energy admitting the Adler-Finch-Skea solution [J. Math. Phys. 15, 727 (1974); Class. Quantum Grav. 6, 467 (1989)] within Einstein gravity. To develop this model, the Karmarkar condition was employed to determine the radial metric component, while Adler's methodology was used to choose the time-metric component. For this purpose, we explored a particular strange star, Her X-1, with observed values of mass $(0.85 \pm 0.15)M_{\odot}$ and radius $= 8.1_{-0.41}^{+0.41}$ km. In this context, we proceeded to model dark energy using the equation of state (EoS), such that the density of dark energy is proportional to the density of isotropic perfect fluid matter. The unknown constants in the metric were determined by smooth matching using the Darmois-Israel criterion. We conduct an in-depth examination of the stability and force equilibrium of our suggested star framework, as well as several physical characteristics of the model such as the metric function, pressure, density, mass-radius relation, and dark energy parameters. Thus, the physical consistency and stability of the present model are investigated. Therefore, following a comprehensive theoretical investigation, we discovered that our proposed model is singularity free and meets all the stability requirements to be a stable and physically realistic stellar model.

\end{abstract}

\maketitle

\section{Introduction}

Over the past few decades, one of the most fascinating discoveries in cosmology has been the realization that our universe is not just expanding-it's doing so at an accelerating rate. This unexpected acceleration has been widely attributed to a mysterious form of energy known as dark energy, which is believed to make up nearly $70 \%$ of the universe. Dark energy is typically modeled as a spatially uniform cosmic fluid. However, this concept can be extended to describe non-uniform, spherically symmetric spacetimes by treating the pressure in its equation of state as a negative radial pressure, with the transverse pressure derived from the Einstein field equations \cite{Lobo:2005uf}. Despite the ongoing expansion of the universe, the influence of dark energy remains constant due to its uniform distribution across space and time. In 1998, the Supernova Cosmology Project at Lawrence Berkeley National Laboratory and the High-z Supernova Search Team observed type-Ia supernovae \cite{SupernovaCosmologyProject:1998vns}, leading to the groundbreaking conclusion that the universe's expansion is accelerating. These supernovae continue to serve as the most compelling observational evidence for the existence of dark energy. While dark energy is usually discussed in a cosmological context, recent studies have started exploring its role in compact astrophysical objects, such as stars. This has led to intriguing models of dark energy stars, which challenge traditional ideas of stellar collapse and black hole formation.\par
A considerable amount of research has focused on compact astrophysical objects whose internal pressure $p$ and energy density $\rho$ follow an equation of state commonly associated with dark energy, such as $p = -\rho$. These objects have been referred to by various names in the literature, but for simplicity, we refer to them here as "dark energy stars" \cite{beltracchi2019formation, Chapline:2004jfp}. Although the exact nature of dark energy remains elusive, its presence can be inferred through observations of the universe's accelerated expansion and the rate at which large-scale cosmic structures-like galaxies and clusters form due to gravitational instabilities.\par
While some models of dark energy stars allow for spacetime singularities, interest generally leans toward configurations that are free from such singularities. Recent astronomical data has confirmed that the universe's expansion is accelerating. Notably, Riess and collaborators, using the Hubble Space Telescope, observed that this expansion is occurring approximately $9\%$ faster than previously estimated \cite{Riess:2019cxk}.\par
Dark energy, despite acting in opposition to gravity, plays a key role in accelerating cosmic expansion and simultaneously suppressing the growth of large-scale structures. This behavior leads to a violation of the strong energy condition, a cornerstone of classical gravitational theory. Because of dark energy's central role in cosmology, researchers are naturally driven to explore its potential manifestations on local astrophysical scales. The dark energy equation of state is often represented as $p = \omega\rho$, where $\omega < -\frac{1}{3}$ \cite{sushkov2005wormholes, bibi2016solution}. In this context, Feng et al. explored various phenomenological models describing the interaction between dark energy and dark matter, using a combined analysis of three independent observational datasets \cite{feng2008observational}.\par
To study such exotic objects, we turn to the foundation of gravitational physics-Einstein's field equations. These equations describe how matter and energy influence the curvature of spacetime. However, solving them is no easy task due to their complexity. One way to make progress is to introduce geometric constraints that simplify the equations while still capturing essential physical features. Among these, the Karmarkar condition \cite{karmarkar1948gravitational} plays a particularly useful role. It ensures that the interior of a star has an "embedding class-I" geometry, meaning it can be neatly embedded in a higher-dimensional flat space. This condition helps reduce the mathematical complexity by linking the gravitational potentials in a precise way.\par
Another helpful tool in constructing stellar models is choosing an appropriate metric or a mathematical description of spacetime inside the star. A well-known and powerful choice is the Adler-Finch-Skea (AFS) solution \cite{Adler:1974dn,finch}, which has been widely used because of its simplicity and physical relevance. Originally proposed for uncharged, isotropic fluid spheres, the AFS metric has been shown to work well in more complex scenarios too. It behaves well at the center of a star and provides finite, realistic values for key physical quantities like pressure and density.\par
In this study, we build on these ideas by exploring a model of a charged dark energy star based on the Adler-Finch-Skea solution, while also satisfying the Karmarkar condition. Including electric charge in stellar models is not just a mathematical curiosity. In extremely compact stars-like neutron stars or hypothetical quark stars-the electric field can be enormously strong. Charge can help stabilize such stars against gravitational collapse and can even influence their overall structure and behavior. Ray and collaborators investigated how the presence of electric charge influences compact stars by assuming that the charge distribution is directly proportional to the mass density. They found that incorporating electric charge leads to a modified form of the relativistic hydrostatic equilibrium equation, commonly known as the Tolman-Oppenheimer-Volkoff (TOV) equation \cite{ray2003electrically}. On the other hand, several specific models of dark energy stars have been introduced in recent literature \cite{chan2009star, chan2009anisotropic, lobo2007gravastars, chan2011gravastars, bertolami2005chaplygin, cattoen2005visser}.\par
The inclusion of dark energy inside the star adds another layer of complexity and interest. By assuming a suitable equation of state for the dark energy component, we can explore how it interacts with normal matter and electromagnetic fields in a stellar setting. Our model therefore combines anisotropic pressure, charge, and dark energy-three key ingredients that allow us to go beyond simple idealized stars and move toward more realistic and diverse compact objects.\par
To construct the model, we solve the Einstein-Maxwell field equations \cite{gupta2011class,kiess2012exact,takisa2013some,malaver2017new,malaver2018generalized,sunzu2014quark} under the assumption of spherical symmetry, with the AFS metric and Karmarkar condition guiding the form of the solution. We then analyze the resulting model using a set of physical checks: Does the energy density stay positive and finite? Are the pressures well-behaved? Is the solution stable against small disturbances? Does it obey causality? And finally, can it be smoothly matched to an exterior spacetime- the Reissner-Nordstr\"om solution, in this case?\par
The goal of this work is to understand how these theoretical tools-AFS metric and Karmarkar condition-can help us model more exotic stars that may exist in nature. By exploring these charged dark energy stars, we may uncover new possibilities for what lies inside the densest regions of the universe. These models may also contribute to interpreting future astronomical observations, especially those involving compact objects with unusual behavior that current models struggle to explain. Several key findings have been highlighted related to the formation of dense and stable stellar structures driven by the repulsive nature of scalar field of dark energy.\par

Several researchers have investigated the characteristics of compact stars, such as the work by Nazar et al.~\cite{Nazar2025-gl} explored the structural characteristics of static, anisotropic compact spheres in the context of non-conservative gravity and reported new insights relevant for realistic modeling of massive stellar configurations. Recent work by Shahzad~\cite{Shahzad:2025fzk} employs the MIT Bag equation of state within the framework of modified gravity \(f(R,T) = R + \zeta T\) to model electrically charged anisotropic compact stars, providing useful theoretical insights applicable to the study of charged dark energy stars and dense stellar configurations; Ditta et al.~\cite{Ditta2025-yc} have explored oscillatory motion and quasi-periodic oscillation (QPO) signatures around rotating compact objects with hair parameters, providing valuable insights into the dynamics around exotic astrophysical bodies. These findings contribute to the understanding of stellar objects with complex structures, relevant for modeling charged dark energy stars; Mustafa et al.~\cite{Mustafa:2020yux} provided a comprehensive study of anisotropic compact stars under the Karmarkar condition within the framework of \(f(R,T)\) gravity. Their work yielded new exact solutions for relativistic stellar models that are free from singularities and satisfy all energy and stability conditions, offering valuable tools for modeling realistic compact stars with anisotropy; Mustafa et al.~\cite{Mustafa:2020jln} investigated physically viable anisotropic compact star models in the framework of \(f(R,G)\) gravity, where \(R\) and \(G\) are the Ricci scalar and Gauss-Bonnet invariant, respectively. They employed the Karmarkar condition to obtain exact solutions that satisfy energy conditions and stability criteria, providing valuable models for realistic compact stars that can be utilized for comparison with observed stellar objects; and further references therein.

This article is structured into five main sections. In Section \ref{interior}, we outline the method used to solve the Einstein-Maxwell field equations for a charged, static, and spherically symmetric matter distribution interacting with dark energy, utilizing the well-known AFS metric. The constants in our model are determined using appropriate junction conditions, specifically applied to the compact star candidate Her X-1. Section \ref{phy} is devoted to examining the physical properties of the proposed model. In Section \ref{stable}, we carry out a detailed stability analysis from multiple perspectives and investigate the force balance within the system. Finally, in Section \ref{con}, we present a summary of our key results along with concluding remarks

\section{Mathematical Foundation of Einstein-Maxwell space-time and Karmakar Condition}\label{interior}

It is convenient to use the line element in canonical form to depict the interior geometry of a static, spherically symmetric, superdense object in a spherical Schwarzschild coordinate system $x^\mu= (t,\,r,\,\theta,\,\phi)$ as
\begin{equation} \label{line1}
ds^{2}_{\textbf{int}}= g_{\mu \nu}dx^\mu dx^\nu = -e^{\eta (r)} dt^2 + e^{\beta (r)} dr^2 + r^2 (d\theta^2 + \sin^2\theta\, d\phi^2),
\end{equation}
Here $\eta$ and $\beta$ are functions of the radial coordinate $'r'$ only. These metric coefficients serve as crucial for interpreting the gravitational environment of a stellar model. The spacetime signature is represented as $(-, +, +, +)$.\par
Assume that matter with matter-energy density $\rho$, pressure $p$ of the corresponding baryonic matter, electric field intensity $E$, and 'dark' energy density $\rho^{D}$, corresponding radial pressure $p_r^{D}$, and tangential pressure $p_t^{D}$, respectively, make up the energy-momentum tensor for an anisotropic charged with two fluids. The sub-index
"D" stands for the dark energy term throughout the work. The (variable) cosmological constant ($\beta$) can be used to express this dark energy density as follows: $\rho^{\text{D}}=\frac{\beta}{8\pi}$ \cite{Ghezzi:2009ct}.\par
For an anisotropic charged system with two fluids, the associated energy-momentum tensor can be written as
 \begin{equation} \label{t1}
  \begin{rcases}
    \begin{aligned}
      \mathcal{T}_0^0 &=\rho^{\text{eff}}+E^2= (\rho+\rho^{D}) +E^2, \\
      \mathcal{T}_1^1 &=-p_r^{\text{eff}}+E^2=-(p+p_r^{D}) +E^2, \\
      \mathcal{T}_2^2 &=\mathcal{T}_3^3=-p_t^{\text{eff}}-E^2=-(p+p_t^{D}) -E^2 ,\\
      \mathcal{T}_0^1 &=\mathcal{T}_1^0=0.
    \end{aligned}
  \end{rcases} \text{Energy-Momentum Tensor}
\end{equation}
where, $\rho^{\text{eff}}=(\rho+\rho^{D}$), $p_r^{\text{eff}}=(p+p_r^{D}$), and $p_t^{\text{eff}}=(p+p_t^{D}$) are the effective energy density and pressure components, respectively.\par
Now, assuming  the relativistic geometrized unit system ($G = c = 1$), the corresponding Einstein-Maxwell field equations are as follows,
\begin{eqnarray}
\mathcal{T}_0^0:~\kappa(\rho+\rho^{D}) +E^2 &=&e^{-\beta}\left[\frac{\beta'}{r}-\frac{1}{r^{2}} \right]+\frac{1}{r^{2}}, \label{fe1}\\
\mathcal{T}_1^1:~\kappa (p+p_r^{D}) -E^2  &=&e^{-\beta}\left[\frac{1}{r^{2}}+\frac{\eta'}{r} \right]-\frac{1}{r^{2}}, \label{fe2}\\
\mathcal{T}_2^2=\mathcal{T}_3^3:~\kappa (p+p_t^{D}) +E^2 &=& \frac{1}{2}e^{-\beta}\left[ \frac{\eta'^{2}}{2}+\eta''-\frac{\beta'\eta'}{2}+\frac{\eta'-\beta'}{r}\right]\label{fe3}
\end{eqnarray}
where the prime symbol $(^{\prime})$ denotes the differentiation with respect to the radial coordinate `r' and $\kappa (=8\pi)$ is Einstein's constant.\par
Here $E (r)$ signifies the electric field intensity at the interior part, while $\sigma=\sigma(r)$ is the charge density connected by the relation,
\begin{eqnarray}
\kappa \sigma r^2 &=& \frac{2 (r^2 E)'}{\sqrt{e^{\beta}}},\label{fe4}
\end{eqnarray}
Now using the relativistic Gauss's law, equation (\ref{fe4}) can also be stated as follows,
\begin{eqnarray}
E r^2 &=& \frac{\kappa}{2} \int_0^r \sigma r^2 \sqrt{e^{\beta}}  dr = q(r),\label{fe5}
\end{eqnarray}
where $q(r)$ represents the total electric charge contained within the fluid sphere of radius $r$.\par
Now, the necessary and sufficient Karmarkar condition \cite{karmarkar1948gravitational} satisfied by space-time (\ref{line1}) in terms of the Riemannian curvature tensor is given as
\begin{equation}
\mathcal{R}_{0303}=\frac{\mathcal{R}_{0101}\cdot \mathcal{R}_{2323}+ \mathcal{R}_{0113}\cdot\mathcal{R}_{0223}}{\mathcal{R}_{1212}} \label{kar}
\end{equation}
provided with $\mathcal{R}_{1212}\neq 0$ \cite{pandey} and  $\mathcal{R}_{abcd}$ are the non-zero components of the
Riemann tensor. The Karmarkar condition allows us to pick one of the metric potentials, which is then utilized to derive the second potential by integration. A spacetime is classified as an embedding class I if it fulfills the Karmarkar condition (Eq.~\ref{kar}). It is a necessary condition but not a self-sufficient requirement while considering the physical viability of the compact celestial object in question. This condition plays an important role in finding solutions to astrophysical models. This condition can be obtained for both charged and uncharged stellar models.\par
Substituting the components of the Riemannian curvature tensor of condition (\ref{kar}) using the line element (\ref{line1}) provides the following differential equation of the form
\begin{equation}
\frac{\beta'\eta'}{e^{\beta}-1}=2\eta''-\eta' (\beta'-\eta'). \label{diff}
\end{equation}
provided $e^{\beta}\neq 1.$\\ Hence, solving equation (\ref{diff}) we derive
\begin{equation}
e^{\beta}=1+ \mathfrak{F}\eta'^{2}e^{\eta} \label{ebet}
\end{equation}
where $\mathfrak{F} \neq 0$ is an arbitrary integration constant.\par
Now we have to solve the Einstein-Maxwell field equations (\ref{fe1})-(\ref{fe4}) with the help of the above expression (\ref{ebet}). Now, to ensure consistency in the system of equations, we take into account an ansatz of the temporal metric coefficient ($g_{tt}$) proposed by Adler \cite{Adler:1974dn},
\begin{equation}
e^{\eta}= \mathfrak{B}(1+\mathfrak{C}r^{2})^{2} \label{eta}
\end{equation}
where $\mathfrak{B}$ and $\mathfrak{C}$ are arbitrary constants.\par
Hence, using the above expression (\ref{eta}) in (\ref{ebet}) we finally derive
\begin{equation}
e^{\beta}=1 + 16 \mathfrak{B} \mathfrak{C}^2 \mathfrak{F} r^2 \label{beta}
\end{equation}
The above metric expression is identical to the well-known Finch-Skea solution \cite{finch}. The Finch-Skea model that satisfies the Karmarkar condition usually simplifies the analysis by focusing on certain metric potentials without an extensive analysis of boundary conditions. The Finch-Skea model can provide plausible solutions. Still, it may not consider the complex physical factors and conditions required for a smooth transition between the interior and exterior environment of a star. This study uses a rigorous framework that complies with Karmarkar's condition and includes detailed boundary conditions, improving the physical validity and applicability of this stellar model to real-world astrophysical phenomena. Later, we will compute the numerical values of these constants ($\mathfrak{B},~ \mathfrak{C},$ and $\mathfrak{F}$) within the metric potential components through a smooth matching of the interior and exterior spacetimes. These metric potential components provide a non-singular stellar model, which will be explained in upcoming sections.\par

With the help of the metric potential expressions given in (\ref{eta}) and (\ref{beta}), the Einstein-Maxwell field equations (\ref{fe1})-(\ref{fe3}) take the following form :
\begin{eqnarray}
\kappa(\rho+\rho^{D}) +E^2 &=&\frac{16 \mathfrak{B} \mathfrak{C}^2 \mathfrak{F} (3 + 16 \mathfrak{B} \mathfrak{C}^2 \mathfrak{F} r^2)}{(1 + 16 \mathfrak{B} \mathfrak{C}^2 \mathfrak{F} r^2)^2},   \label{f1}\\
\kappa (p+p_r^{D}) - E^2 &=& \frac{\frac{1}{r^2}+\frac{4 \mathfrak{C}}{1 + \mathfrak{C} r^2}}{1 + 16 \mathfrak{B} \mathfrak{C}^2 \mathfrak{F} r^2}-\frac{1}{r^2}, \label{f2}\\
\kappa (p+p_t^{D}) +E^2 &=&\frac{4 \mathfrak{C} (1 + 4 \mathfrak{B} \mathfrak{C} \mathfrak{F} (-1 + \mathfrak{C} r^2))}{(1 + \mathfrak{C} r^2)(1 + 16 \mathfrak{B} \mathfrak{C}^2 \mathfrak{F} r^2)^2} .\label{f3}
\end{eqnarray}

\subsection{Dark energy EoS}\label{eosde}

Obtaining explicit solutions to the aforementioned field equations (\ref{f1})-(\ref{f3}) is too difficult due to its non-linearity nature, despite the system of equations being explicitly mentioned. So, to remove this complexity, we have adopted three hypotheses as proposed earlier by the authors mentioned in Refs. \cite{Ghezzi:2005iy,Ghezzi:2009ct,Barreto:2006cr} given as follows:\\ \\
(i) Radial dark pressure ($p_r^{D}$) is related to dark density through this relation
\begin{eqnarray}\label{as1}
p_r^{D} + \rho^{D} =0,
\end{eqnarray}
(ii) The dark density ($\rho^{D}$) is proportional to the normal ordinary matter density, i.e.,
\begin{eqnarray}\label{as2}
\rho^{D}=\chi \rho,
\end{eqnarray}
where $\chi$ is a non-zero constant and its value will be derived later from the boundary conditions.\\ \\
(iii) Finally, following the prior research work in Refs. \cite{Barreto:2006cr, das2015anisotropic} we preferably assume the difference between radial and tangential dark pressure is proportional to the square of the electric field intensity $E$, connected by the relation,
\begin{eqnarray}\label{as3}
 4\pi\left(p_t^{D} - p_r^{D}\right) = E^2  
\end{eqnarray}

\subsection{Proposed dark energy stellar model}

Now, solving equations (\ref{f1})-(\ref{f3}) with the above hypotheses (\ref{as1})-(\ref{as3}), we derive explicit expressions for $\rho$, $p$ and $E^2$ as follows:
\begin{eqnarray}
 \rho &=& \frac{\mathfrak{B} \mathfrak{C}^2 F (6 + \mathfrak{C} r^2 (7 + 24 \mathfrak{B} \mathfrak{C} \mathfrak{F} (1 + \mathfrak{C} r^2)))}{\pi(1 + \chi) (1 + \mathfrak{C} r^2) (1 + 16 \mathfrak{B} \mathfrak{C}^2 \mathfrak{F} r^2)^2} ,   \label{f11}\\
 p &=&  \frac{\mathfrak{C} \Bigg[1 + \chi \Big(1 + 8 \mathfrak{B} \mathfrak{C} \mathfrak{F} (1 + 3 \mathfrak{C} r^2)\Big) - 
   2 \mathfrak{B} \mathfrak{C} \mathfrak{F} \Big\{2 + \mathfrak{C} r^2 \Big(-5 + 24 \mathfrak{B} \mathfrak{C} \mathfrak{F} (1 + \mathfrak{C} r^2)\Big)\Big\}\Bigg]}{2\pi(1 + \chi) (1 + \mathfrak{C} r^2) (1 + 16 \mathfrak{B} \mathfrak{C}^2 \mathfrak{F} r^2)^2}   ,\label{f12}\\
E^2 &=&  \frac{8 \mathfrak{B}\mathfrak{C}^3 \mathfrak{F} r^2 \Big(-1 + 8 \mathfrak{B} \mathfrak{C} \mathfrak{F} (1 + \mathfrak{C} r^2)\Big)}{(1 + \mathfrak{C} r^2) (1 + 16 \mathfrak{B} \mathfrak{C}^2 \mathfrak{F} r^2)^2} .\label{f13}
\end{eqnarray}
\\ The explicit expressions for matter-energy density, radial and transverse pressure due to the dark energy are obtained as follows:
\begin{eqnarray} 
\rho^{D} &=&  \frac{\chi \mathfrak{B} \mathfrak{C}^2  \mathfrak{F} \Big[6 +  \mathfrak{C} r^2 \Big\{7 + 24 \mathfrak{B} \mathfrak{C} \mathfrak{F} (1 +  \mathfrak{C} r^2)\Big\}\Big]}{\pi(1 + \chi) (1 + \mathfrak{C} r^2) (1 + 16 \mathfrak{B} \mathfrak{C}^2 \mathfrak{F} r^2)^2},    \label{de1}\\
p_r^{D} &=& - \frac{\chi \mathfrak{B} \mathfrak{C}^2  \mathfrak{F} \Big[6 +  \mathfrak{C} r^2 \Big\{7 + 24 \mathfrak{B} \mathfrak{C} \mathfrak{F} (1 +  \mathfrak{C} r^2)\Big\}\Big]}{\pi(1 + \chi) (1 + \mathfrak{C} r^2) (1 + 16 \mathfrak{B} \mathfrak{C}^2 \mathfrak{F} r^2)^2},   \label{de2}\\
p_t^{D}  &=&  \frac{\mathfrak{B} \mathfrak{C}^2 \mathfrak{F} \Big[-6 \chi - C \Big\{2 + 9 \chi + 8 (-2 + \chi) \mathfrak{B} \mathfrak{C} \mathfrak{F}\Big\} r^2 - 
   8 (-2 + \chi) \mathfrak{B} \mathfrak{C}^3 \mathfrak{F} r^4\Big]}{\pi(1 + \chi) (1 + \mathfrak{C} r^2) (1 + 16 \mathfrak{B} \mathfrak{C}^2 \mathfrak{F} r^2)^2}.  \label{de3}
\end{eqnarray}

\subsection{Exterior Geometry and Boundary Conditions}\label{match}

For further analysis of model parameters, the values of $\mathfrak{B}$, $\mathfrak{C}$, and $\mathfrak{F}$ must be determined. To determine the numerical values of unknown constants in the radial and geometric components of the metric variables, we match the interior space-time smoothly at the surface of the strange stellar object to the exterior Reissner-Nordstr\"om space-time \cite{reissner1916eigengravitation,nordstrom1918energy} characterized by the following line element,
\begin{eqnarray}
ds_{\textbf{ext}}^{2} &=& -\left(1 - \frac{2\mathbb{M}}{r} + \frac {\mathbb{Q}^2}{r^2}\right)dt^2 + \frac{dr^2}{\left(1 - \frac{2\mathbb{M}}{r} + \frac {\mathbb{Q}^2}{r^2}\right)}
 + r^2(d\theta^2+\sin^2\theta d\phi^2), \label{eq22}
\end{eqnarray}
where $\mathbb{M}$ is the total mass enclosed within the fluid sphere of radius $r = \mathbb{R}$ and $\mathbb{Q}$ is the total charge enclosed within the surface $r = \mathbb{R}$. This approach leads to a smooth transition over the stellar boundary, ensuring physical consistency between the two regions.
Now imposing the continuity requirement of the metric coefficients $g_{tt}$, $g_{rr}$ and $\frac{\partial g_{tt}}{\partial r}$ across the boundary surface $\Sigma r= \mathbb{R}$ between the interior and the exterior regions, i.e. $[ds_{\textbf{int}}]=[ds_{\textbf{ext}}]$, explicitly produce the following set of relations:
\begin{eqnarray}
1 - 2\tilde{\mathbb{U}}+ \tilde{\mathbb{V}} = e^{\eta (\mathbb{R})} &=& \mathfrak{B}(1+\mathfrak{C}\mathbb{R}^{2})^{2},\label{eq23}\\
1 - 2\tilde{\mathbb{U}} + \tilde{\mathbb{V}} = e^{-\beta (\mathbb{R})} &=& 1 + 16 \mathfrak{B} \mathfrak{C}^2 \mathfrak{F} \mathbb{R}^2,\label{eq24}\\
\tilde{\mathbb{U}} - \tilde{\mathbb{V}} = \frac{\mathbb{R}}{2}\Bigg\{ \frac{\partial e^{\eta (r)}}{\partial r}  \Bigg\}_{r=\mathbb{R}} &=& 2\mathfrak{B}\mathfrak{C}\mathbb{R}^{2}(1+\mathfrak{C}\mathbb{R}^{2}).\label{eq25}
\end{eqnarray}
where, $\tilde{\mathbb{U}}=\frac{\mathbb{M}}{\mathbb{R}}$ and $\tilde{\mathbb{V}}=\frac {\mathbb{Q}^2}{\mathbb{R}^2} $ and both $\tilde{\mathbb{U}},\,\tilde{\mathbb{V}}$ are dimensionless quantities.\par
Thus the Eqs.~(\ref{eq23})-(\ref{eq25})have been solved to determine the values of the constants $\mathfrak{B}$, $\mathfrak{C}$, and $\mathfrak{F}$ as,
\begin{eqnarray}
\mathfrak{B} &=& \frac{\Big[3 \mathbb{Q}^2 + \mathbb{R} (-5 \mathbb{M} + 2 \mathbb{R})\Big]^2}{4 \mathbb{R}^2 \Big[\mathbb{Q}^2 + \mathbb{R} (-2 \mathbb{M} + \mathbb{R})\Big]},\label{eq26}\\ 
\mathfrak{C} &=& \frac{-\mathbb{Q}^2 + \mathbb{M} \mathbb{R}}{\mathbb{R}^2 \Big[3 \mathbb{Q}^2 + \mathbb{R} (-5 \mathbb{M} + 2 \mathbb{R})\Big]}, \label{eq27}\\ 
\mathfrak{F} &=& \frac{-\mathbb{Q}^2 + 2 \mathbb{M} \mathbb{R}}{16 \mathfrak{B} \mathfrak{C}^2 \mathbb{R}^2 \Big[\mathbb{Q}^2 + \mathbb{R} (-2 \mathbb{M} + \mathbb{R})\Big]} \label{eq28}
\end{eqnarray}
Now, from the continuity condition of the extrinsic curvature at the surface $r=\mathbb{R}$, that implies $p(r=\mathbb{R})=0$, the value of $\chi$ has been determined as follows:
\begin{eqnarray}\label{coup}
\chi &=&\frac{-1 + 2 \mathfrak{B} \mathfrak{C} \mathfrak{F} \Big[2 + \mathfrak{C} \mathbb{R}^2 \Big(-5 + 24 \mathfrak{B} \mathfrak{C} \mathfrak{F} (1 + \mathfrak{C} \mathbb{R}^2)\Big)\Big]}{1 + 8 \mathfrak{B} \mathfrak{C} \mathfrak{F} (1 + 3 \mathfrak{C} \mathbb{R}^2)}
\end{eqnarray}
Thus, we finally obtain all the unknown constants in the metric potentials in terms of $\mathbb{M}, \mathbb{R}$, and $\mathbb{Q}$. Using the expressions mentioned above, we can derive the numerical values of the model parameters for compact stars in our universe. Now, to properly describe the physical characteristics of our present model, we have considered here, particularly the compact stellar candidate Her X-1 with observed mass and radius $\mathbb{M} = 0.85 \pm 0.15~M_{\odot},\, \mathbb{R} = 8.1_{-0.41}^{+0.41}$ km \cite{Abubekerov:2008inw}. In addition to this, we have also assumed $\mathbb{Q} = 1.31$ throughout this work. The motivation for this specific selection stems from the findings of Varela et al.~\cite{Varela:2010mf}, who established that the square of the charge-to-radius ratio, $\big(\frac{\mathbb{Q}^2}{\mathbb{R}^2}\big)$, for a charged anisotropic fluid sphere must remain within the interval $[0, 0.543)$. For the selected compact object HER X-1, the computed value $\frac{\mathbb{Q}^2}{\mathbb{R}^2} = 0.0237522$ (approximately) falls well within the prescribed interval. This compliance reinforces the accuracy of our numerical results and further supports the reliability of the graphical analysis. Furthermore, in Table~\ref{table111}, the numerically computed values of $\mathfrak{B}$, $\mathfrak{C}$, $\mathfrak{F}$, and $\chi$ have been presented for some well-known compact star candidates.

\begin{table*}[t]
\centering
\caption{The numerical values of $\mathfrak{B},\,\mathfrak{C}$, $\mathfrak{F}$, and $\chi$ for some well-known stellar candidates (Assuming $\mathbb{Q} = 1.31$).}\label{table111}
\begin{tabular}{@{}ccccccccccccccc@{}}
\hline
Star & Observed mass & Observed radius & Estimated  & Estimated &  $\mathfrak{B}$&$\mathfrak{C}$&$\mathfrak{F}$ & $\chi$\\
& $M_{\odot}$ & km. & mass ($M_{\odot}$) & radius (km.)&  & $km^{-2}$ &  $km^{2}$&\\
\hline
Her X-1 \cite{Abubekerov:2012yj}& $0.85 \pm 0.15$ & $8.1 \pm 0.41$ & 0.85 & 8.5 &0.749269  &0.000671648 &547.58     &0.0436839   \\
SMC X-4 \cite{Rawls:2011jw} & $1.29 \pm 0.05$ & $8.831 \pm 0.09$ & 1.29 & 8.8 &0.609858  &0.00120492  &338.887  &0.0189877\\
Vela X-1 \cite{Rawls:2011jw} & $1.77 \pm 0.08$ & $9.56 \pm 0.08$    & 1.77 & 9.5 & 0.489908  &0.00164709   &285.052       & 0.00321299\\
4U 1538-52 \cite{Rawls:2011jw}& $0.87 \pm 0.07$ & $7.866 \pm 0.21$ & 0.87 & 7.8 &0.723954   &0.000897021 &426.829   & 0.0473986\\
LMC X-4 \cite{Rawls:2011jw} & $1.04 \pm 0.09$ & $8.301 \pm 0.2$ & 1.04 & 8.3 &0.677172  &0.00100624   &385.68    &0.0324208\\
Cen X-3 \cite{Rawls:2011jw} & $1.49 \pm 0.08$ & $9.178 \pm 0.13$ & 1.49 & 9.2&0.561888  &0.00133803  &320.071   &0.011301\\
PSR J1903+327 \cite{Freire:2010tf} & $1.667 \pm 0.021$ & $9.438 \pm 0.03$ & 1.67 & 9.4 &0.515291  &0.00153067  &296.334             &0.00590445\\
4U 1820-30 \cite{Guver:2008gc} & $1.58 \pm 0.06$&  $9.316 \pm 0.086$  & 1.58 & 9.3&0.538283 &0.00143372 &307.257  &0.00850393\\
EXO 1785-248 \cite{Ozel:2008kb} & $1.3 \pm 0.2$ & $8.849 \pm 0.4 $ &1.4 &8.85 &0.575826 &0.00136495 &310.456  &0.0149181\\
\hline
\end{tabular}
\end{table*}

\section{Physical Analysis of proposed stellar model}\label{phy}

To develop accurate models of stellar structures, such as neutron stars, ultracompact stars, and quark stars, requires a thorough grasp of the structural characteristics of gravitationally confined configurations. Together, pressure, matter density, anisotropy, and the behavior of geometric variables determine a star configuration's stability and equilibrium under its own gravity. The physical attributes of the proposed stellar model are visually analyzed in this section. We may learn more about the internal structure and behavior of the models by plotting several physical parameters as density, pressure, electric field intensity, etc. We can study important aspects such as the evolution of density and pressure from the core to the surface and the function of the electric field in preserving stability from these graphical representations. This investigation clarifies the models' advantages and disadvantages by confirming that they satisfy fundamental physical requirements like central
regularity and realistic behavior throughout the star's radial profile.

\subsection{Nature of metric potentials}\label{mp}

\begin{figure}[H]
    \centering
        \includegraphics[scale=.55]{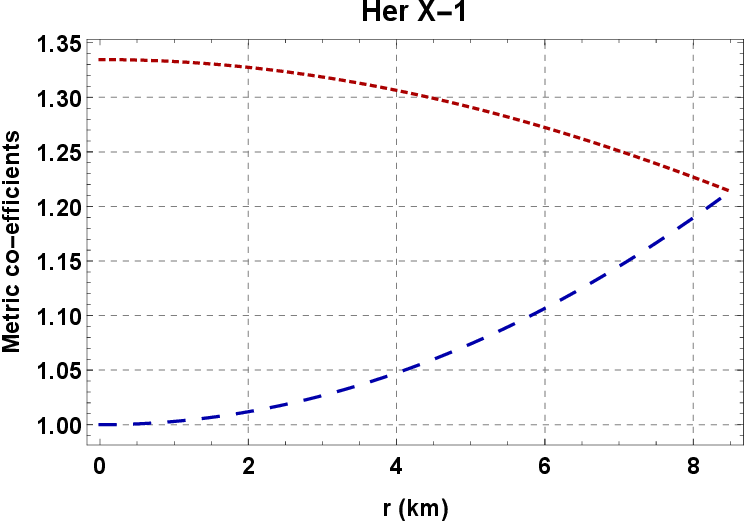}
        \caption{Variation of metric functions $e^{\beta(r)}$ and $e^{-\eta(r)}$ with respect to `r' with required magnified inset.}\label{metric}
\end{figure}

In this subsection, we have focused on analyzing the behavior of the temporal components of the metric potential $e^{\eta(r)}$ and the spatial components $e^{\beta(r)}$. We guarantee that these potentials remain continuous and without any singularities inside the star by selectively choosing them.
We can easily check that $[{e^{\eta(r)}}]_{r = 0}= \mathfrak{B}$, a non-zero constant, and $[{e^{-\beta(r)}}]_{r=0} = 1$, confirming the fact that both metric potential components are finite at the center and have regularity at all points within $r < \mathbb{R}$ \cite{Delgaty:1998uy, Pant:2010iub}. Moreover, $\Big[\frac{d(e^{\eta(r)})}{dr}\Big]_{r=0} = 4 \mathfrak{B} \mathfrak{C} r (1 + \mathfrak{C} r^2)\Big\rvert_{r=0} =0$ \text{and}\\ $\Big[\frac{d(e^{\beta(r)})}{dr}\Big]_{r=0} = 32 \mathfrak{B} \mathfrak{C}^2 \mathfrak{F} r\Big\rvert_{r=0} =0$,\par

i.e., the derivatives of the metric potential components vanish at the star's center. These metric potentials are smooth, positive, and monotonically increasing within the stellar interior, which can be easily verified from the radial profiles of the metric coefficient components shown in Fig.~(\ref{metric}). The internal space-time smoothly matches with the asymptotically flat exterior space-time on the surface ($r = \mathbb{R}$), fulfilling the Darmois-Israel condition \cite{Chu:2021uec, darmois1927equations, Israel:1966rt}. By matching those, we obtain the values of the unknown constants that characterize this model. Thus, we verify that the gravitational potentials are physically well-behaved within the range $(0, \mathbb{R})$.

\subsection{Evolution of ordinary baryonic matter-energy density, pressure, and electric field}

\begin{figure}[H]
    \centering
        \includegraphics[scale=.45]{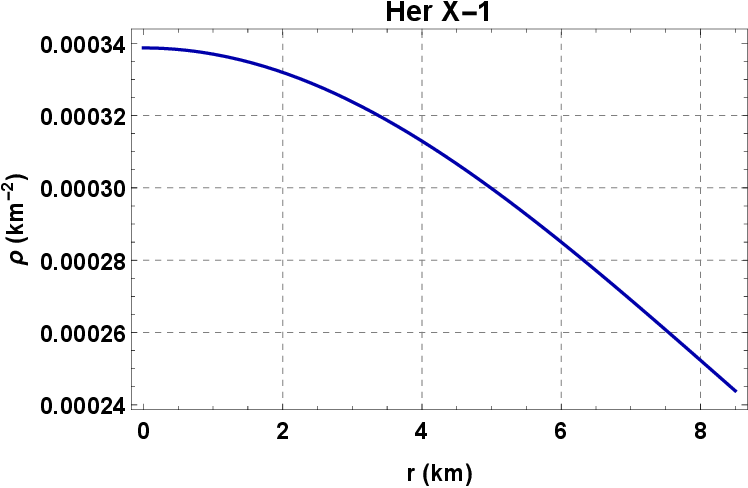}
         \includegraphics[scale=.45]{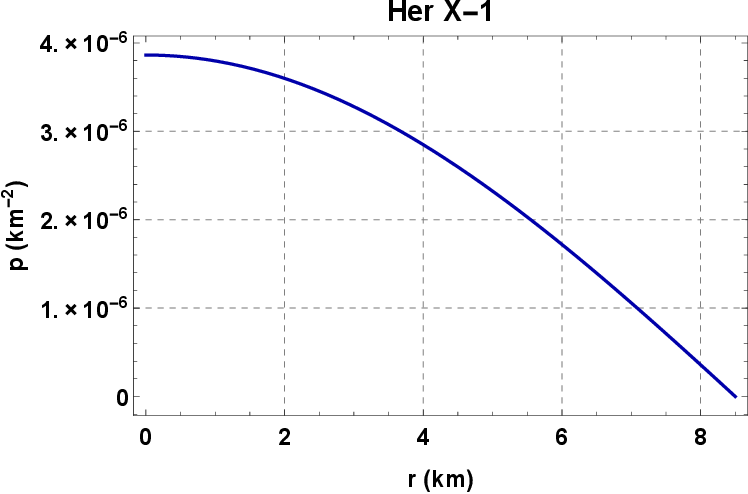}
         \includegraphics[scale=.45]{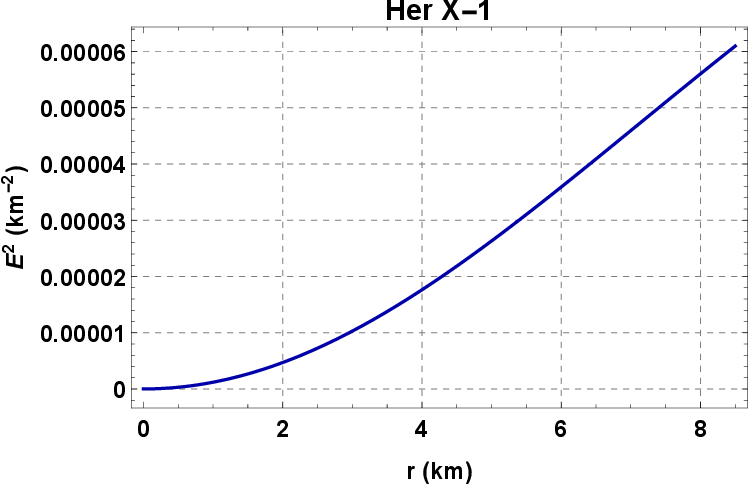}
        \caption{Profiles of baryonic matter-energy density, pressure, and electric field intensity ($E^2$) with respect to `r'.}\label{rho}
\end{figure}

\begin{figure}[htbp]
    \centering
        \includegraphics[scale=.55]{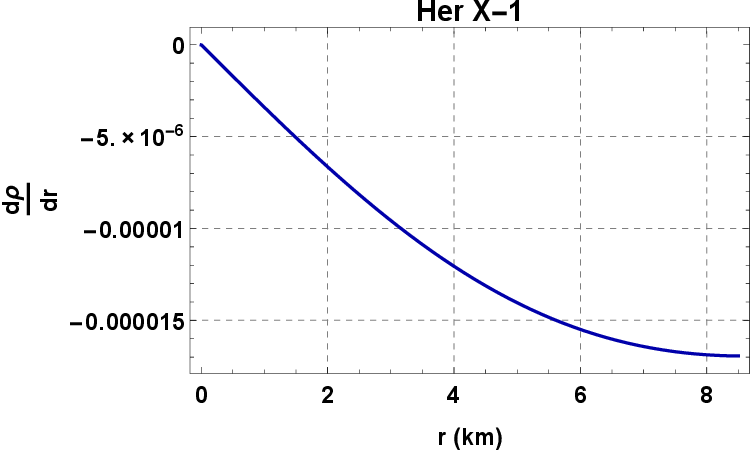}
         \includegraphics[scale=.55]{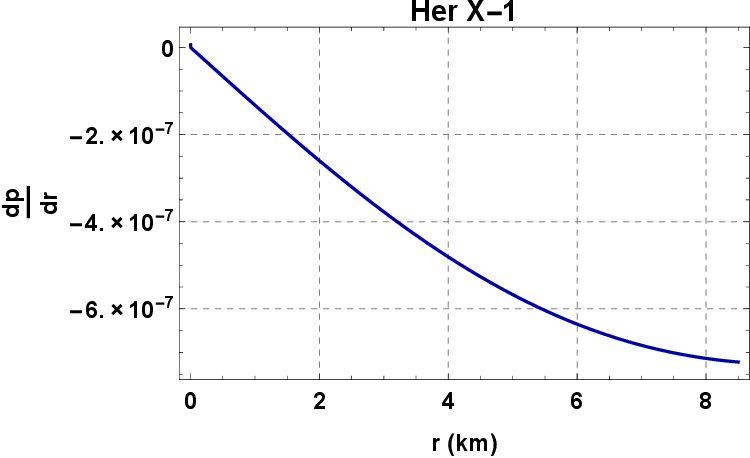}
        \caption{Gradients of matter density and pressure are plotted `r'.}\label{grad}
\end{figure}

\begin{figure}[htbp]
    \centering
        \includegraphics[scale=.55]{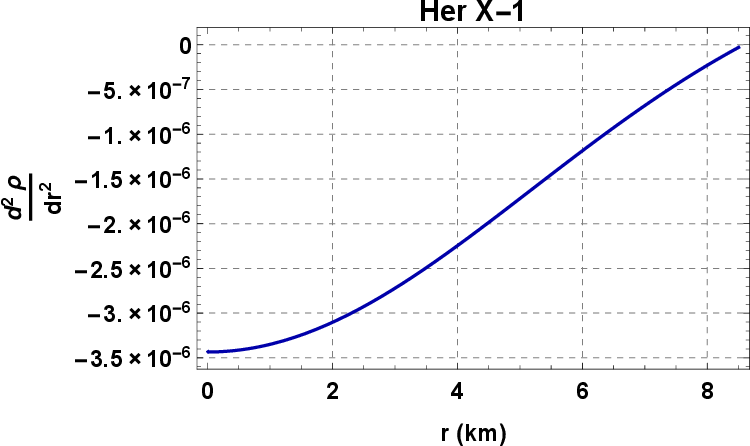}
         \includegraphics[scale=.55]{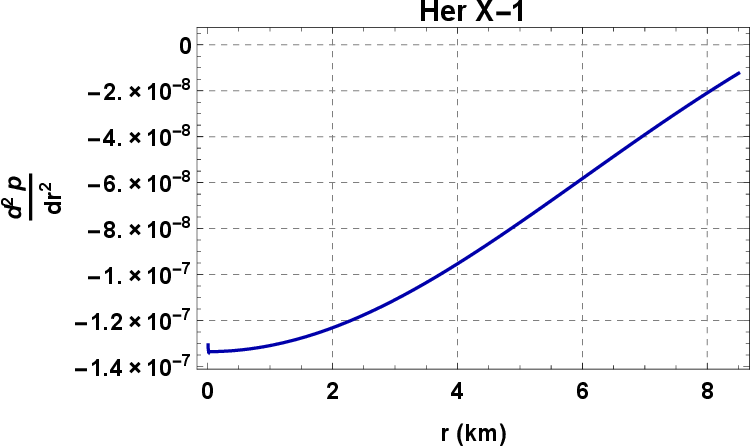}
        \caption{Graphical representation of $\frac{d^2\rho}{dr^2}$ and $\frac{d^2 p}{dr^2}$ versus areal radius `r'.}\label{ddr}
\end{figure}

The density of the confined matter has a major part in determining how stable a stellar structure is against gravitational collapse. Similarly, pressure plays a significant role in establishing the stellar border and overall stability \cite{chandrasekhar1984stars}. For $\rho~ \text{and}~ p$, we obtain the central values from equations (\ref{f11}) and (\ref{f12}) as follows:
$\rho_c= [\rho]_{r=0} = \frac{6 \mathfrak{B} \mathfrak{C}^2 \mathfrak{F}}{\pi + \chi \pi} > 0$,
 $p_c= [p]_{r=0} =  \frac{\mathfrak{C} \Big[1 + \chi + 4 (-1 + 2 \chi) \mathfrak{B} \mathfrak{C} \mathfrak{F}\Big] }{2\pi(1 + \chi)} >0$. 
 Matter-energy density and pressure are monotonically decreasing functions of radius $r$, with a maximum value near the star's center, as seen in Fig.~(\ref{rho}).  $[E^2]_{r=0} = 0.$, meaning that the intensity of the electric field diminishes in the center and progressively increases as $r$ increases, as illustrated in Fig.~(\ref{rho}). Finally, the non-negative values of $\rho$, $p$, and $E^2$ inside the star are also observed. The central density ($\rho_c$), surface density ($\rho_s$), central pressure ($p_c$), and the central pressure-density ratio $p_c/\rho_c$ have all been numerically represented in Table~\ref{tab1}.
 Also, for our current model, the ordinary baryonic matter's density and pressure gradients are determined as follows:
\begin{eqnarray}
  \frac{d\rho}{dr} &=& -\frac{2 \mathfrak{B} \mathfrak{C}^3 \mathfrak{F} r \Bigg[1 + 
   8 \mathfrak{B} \mathfrak{C} \mathfrak{F} \Big[21 + \mathfrak{C} r^2 \Big\{44 + 25 \mathfrak{C} r^2 + 48 \mathfrak{B} \mathfrak{C} \mathfrak{F} (1 + \mathfrak{C} r^2)^2\Big\}\Big]\Bigg]}{\pi (1 + \chi) (1 + \mathfrak{C} r^2)^2 (1 + 16 \mathfrak{B} \mathfrak{C}^2 \mathfrak{F} r^2)^3}, 
\end{eqnarray}
\begin{eqnarray}
 \frac{dp}{dr} &=& -\frac{1}{(1 + \chi) \pi (1 + \mathfrak{C} r^2)^2 (1 + 16 \mathfrak{B} \mathfrak{C}^2 \mathfrak{F} r^2)^3} \Bigg[C^2 r \bigg[1 + \chi (1 + 16 \mathfrak{B} \mathfrak{C} \mathfrak{F} (1 + 3 C r^2 + 16 \mathfrak{B} \mathfrak{C} \mathfrak{F} (1 +  \\&& 3 \mathfrak{C} r^2 (1 + \mathfrak{C} r^2)))) + 2 \mathfrak{B} \mathfrak{C} \mathfrak{F} (9 + 8 \mathfrak{C} (3 r^2 - \mathfrak{B} \mathfrak{F} (5 + \mathfrak{C} r^2 (-4 - 23 \mathfrak{C} r^2 + 48 \mathfrak{B} \mathfrak{C} \mathfrak{F} (1 + \mathfrak{C} r^2)^2))))\bigg]\Bigg]
\end{eqnarray}
In Fig.~(\ref{grad}), we can easily check that the ordinary matter density and pressure gradients remain negative throughout the compact object and vanish at the center.
Now, taking the second derivative, $\rho, p$ we get,
$$\Big[\frac{d^2\rho}{dr^2}\Big]_{r=0} = -\frac{2 \mathfrak{B} \mathfrak{C}^3 \mathfrak{F} (168 \mathfrak{B} \mathfrak{C} \mathfrak{F}-1)}{\pi (1 + \chi)},$$\\
$$\Big[\frac{d^2 p}{dr^2}\Big]_{r=0} = -\frac{\mathfrak{C}^2 \Bigg[1 + 2 \mathfrak{B} \mathfrak{C} \mathfrak{F} (9 - 40 \mathfrak{B} \mathfrak{C} \mathfrak{F}) + 
   \chi \Big[1 + 16 \mathfrak{B} \mathfrak{C} \mathfrak{F} (1 + 16 \mathfrak{B} \mathfrak{C} \mathfrak{F})\Big]\Bigg]}{\pi (1 + \chi)}$$
We can easily check from further graphical investigation from Fig.~(\ref{ddr}) that the second order derivatives assume negative values in the center, implying that both matter density and pressure attain their maximum values at the core of the stellar configuration.

\subsection{Dark energy density and dark pressure}

In physical cosmology, dark energy is a hypothetical form of energy that occupies most of the space and tends to accelerate expansion of the universe \cite{Peebles:2002gy}. It is referred to as "dark" because it does not possess an electric charge and does not react with light or other electromagnetic radiation. Dark energy has the opposite effect of positive energy: it accelerates the expansion of the universe. Dark energy presently accounts for more than three-quarters of the universe's total mass energy, according to the standard cosmological model. Dark energy is being used to build a cyclic model of the cosmos \cite{Baum:2006ee}. Dark energy, unlike other forces, causes the universe to expand due to its extremely negative pressure.  The negative pressure causes gravitational repulsion.\par
\begin{figure}[H]
    \centering
        \includegraphics[scale=.46]{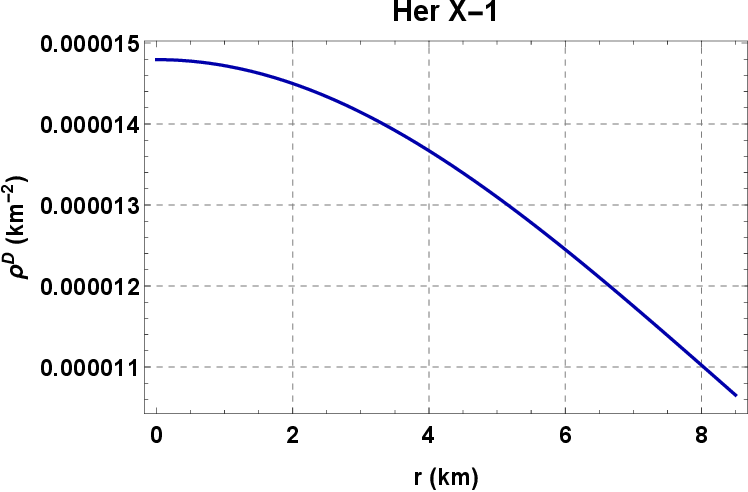}
        \includegraphics[scale=.46]{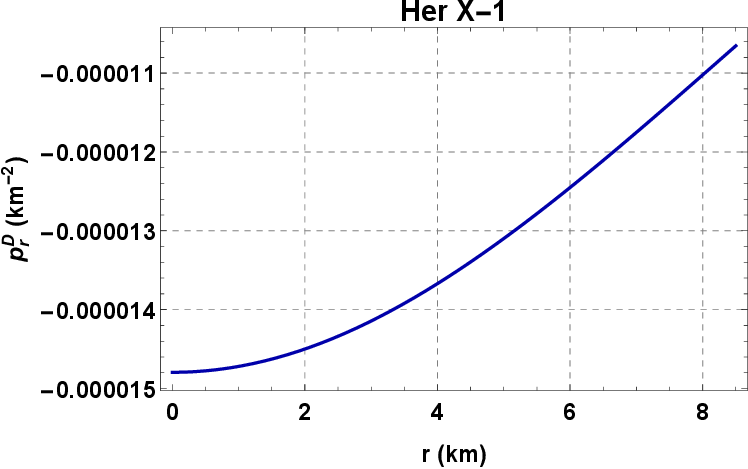}
         \includegraphics[scale=.46]{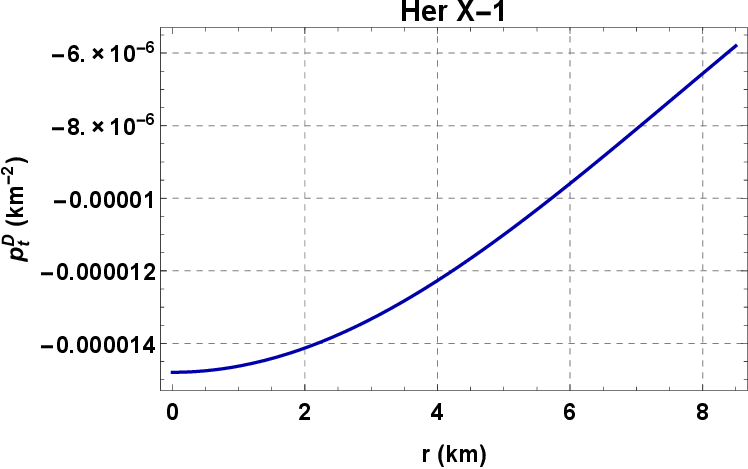}
        \caption{The variation of dark energy density and dark pressure components with respect to `r'.}\label{dark}
\end{figure}

This work simulates the repulsive nature of dark energy, resulting in a negative radial dark pressure ($p_r^{D}<0$) and a positive energy density ($\rho^{D}>0$). Fig.~(\ref{dark}) depicts the profiles of dark energy density, radial dark pressure, and transverse dark pressure. The dark energy density steadily declines within the stellar structure, with the maximum value occurring at the core. Equation (\ref{de1}) yields the maximum value, $\rho^{D}_{max} = [\rho^{D}]. _{r=0} = \frac{6 \chi \mathfrak{B} \mathfrak{C}^2 \mathfrak{F}}{\pi + \chi \pi}$. For this model, it is also interesting that at the center $(r = 0)$,
$p^{D}_r = p^{D}_t = - \rho^{D}_{max}$.

\subsection{Mass-radius relationship and compactness factor}

\begin{figure}[H]
    \centering
        \includegraphics[scale=.5]{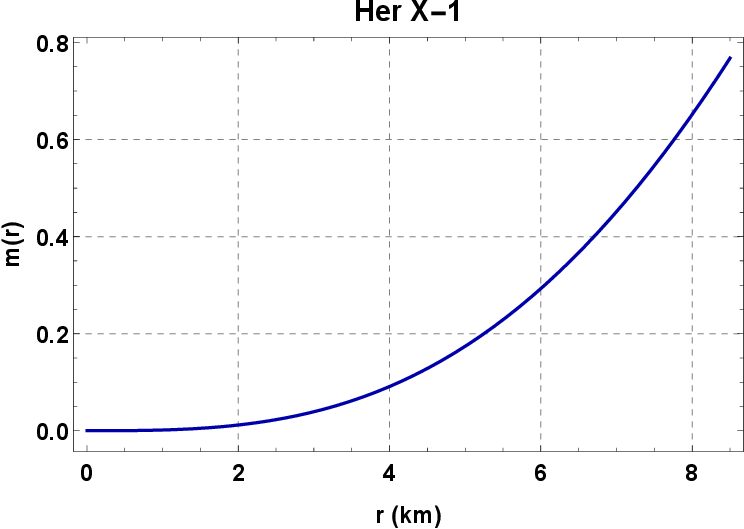}
        \includegraphics[scale=.5]{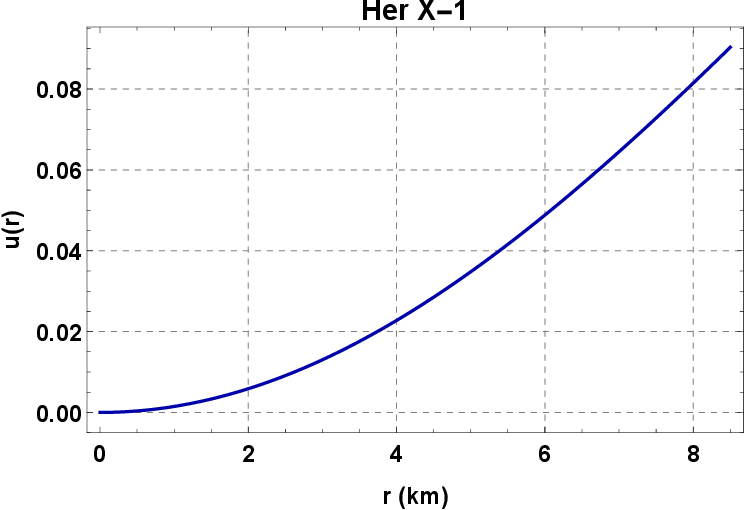}
       \caption{The mass function and compactness factor are plotted against `r'.}\label{mass}
\end{figure}

The mass function $\mathfrak{m}(r)$ and compactness factor $\mathfrak{u}(r)$ are considered the fundamental aspects of compact structural analysis. Since it is gravitationally confined to a finite spatial extent $(r = \mathbb{R})$, we know that the active mass depends on the density profile and grows with the confining radius \cite{buchdahl1959general, glendenning2012compact}. We can simply estimate the mass function $\mathfrak{m}(r)(r)$ for an electrically charged fluid sphere by evaluating the integral directly related to the energy density (\ref{f11}) using the expression \cite{florides1983complete, kumar2022isotropic},
\begin{eqnarray}
 \mathfrak{m}(r) &=& 4\pi\int^r_0{\Big[(\rho + \rho^D) r^2+ r \sigma q e^{\beta/2}\Big]\,dr} \nonumber\\
 &=& \frac{4 \mathfrak{B} \mathfrak{C}^2 \mathfrak{F} r^3 \Big[2 + \mathfrak{C} r^2 \big\{1 + 40 \mathfrak{B} \mathfrak{C} \mathfrak{F} (1 + \mathfrak{C} r^2)\big\}\Big]}{(1 + \mathfrak{C} r^2) (1 + 16 \mathfrak{B} \mathfrak{C}^2 \mathfrak{F} r^2)^2},~~~\label{mm}
 \end{eqnarray}
 It should be clarified that the mass function $\mathfrak{m}(r)$ is a function of $r$ such that $\mathfrak{m}(r = 0) = 0$ and $\mathfrak{m}(r = \mathbb{R}) = \mathbb{M}$. In Fig.~(\ref{mass}), the profile of the mass function (\ref{mm}) has been shown against $r$. Since the mass is directly proportional to the radial distance $r$, it is evident that the mass is regular in the center. As shown in Fig.~(\ref{mass}), maximum mass is attained at the surface $r=\mathbb{R}$.
Moreover, the compactness factor is also defined by a dimensionless parameter $\mathfrak{u}(r) = \frac{\mathfrak{m}(r)}{r}$. The compactness factor $\mathfrak{u}(r)$ is crucial in determining the gravitational properties and stability of dense objects. Buchdahl \cite{Buchdahl:1959zz} demonstrated that $\mathfrak{u}(r)$ has an upper limit of 4/9, preventing gravitational collapse. Fig.~(\ref{mass}) depicts the visual representation of $\mathfrak{u}(r)$, which shows a monotonically increasing behavior versus `r' and does not cross the Buchdahl limit.

\subsection{Surface redshift and Gravitational redshift}

\begin{figure}[H]
    \centering
        \includegraphics[scale=.52]{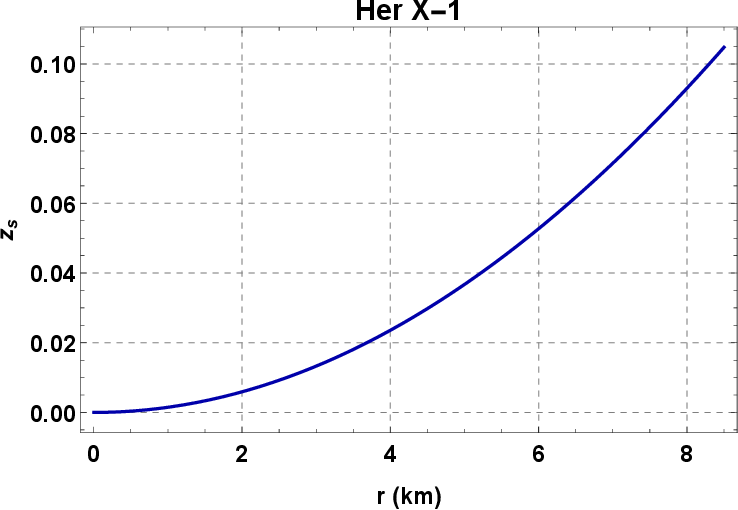}
        \includegraphics[scale=.52]{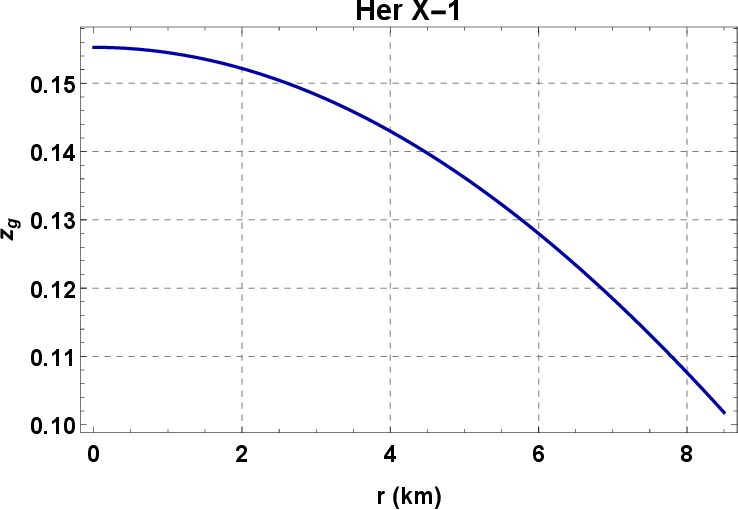}
         \caption{Surface redshift and gravitational redshift are plotted against `r' with required magnified inset.}\label{red}
\end{figure}

The surface redshift $z_s(r)$ for this present model is calculated using the compactness factor $\mathfrak{u}(r)$ expression, which is $z_s(r)=\frac{1}{\sqrt{1-2\mathfrak{u}(r)}}-1$. Fig.~(\ref{red}) shows the variation variation of $z_s(r)$ from center to surface for the compact object chosen. The surface redshift is determined by the stellar mass and radius, also known as surface gravity. In Table~\ref{tab1}, the numerical results of the compactness factor ($\mathfrak{u}(\mathbb{R})$) and surface redshift ($z_s(\mathbb{R})$) for different known compact objects.
However, $z_g(r)=e^{-\frac{\eta(r)}{2}}-1$ indicates the gravitational redshift (or internal redshift) $z_g(r)$ within a static line element. We observe that $z_g(r)$ reaches its maximum in the center and then progressively falls as the radius increases to reach its minimum at the surface. In Fig.~(\ref{red}), we illustrated the variation of $z_g(r)$.
These plots clearly show that the surface redshift and gravitational redshift have completely opposing tendencies throughout this model and also have no singularity throughout their configuration.

\begin{table}
\begin{center}
\caption{\label{tab1} The numerically obtained values of central density $(\rho_c)$, surface density $(\rho_s)$, central pressure $(p_c)$, compactness ratio $(\mathfrak{u})$, surface redshift $(z_s)$, and $p_c/\rho_c$ have been shown for different known compact objects.}
\begin{tabular}{cccccccc}
\hline
\hline  
$\rm{Star}$  & $\rho_c$  &   $\rho_s$  &   $p_c$  &       $u(\mathbb{R})$      &    $z_s(\mathbb{R})$  &   $p_c/\rho_c$  \\
  &  $\text{gm}~\text{cm}^{-3}$  &   $\text{gm}~\text{cm}^{-3}$  &$\text{dyne}~\text{cm}^{-2}$  &     &     &     \\  
\hline
     Her X-1 & $  4.57002\times 10^{14}$  & $ 3.29086\times 10^{14}$  & $ 5.21278\times 10^{12}$ &0.090327  &0.104756  &0.0114065   \\

     SMC X-4 & $ 7.58844\times 10^{14}$  & $ 4.46474\times 10^{14}$  & $1.54175 \times 10^{13}$ &0.140477  & 0.179293 &0.020317    \\

     Vela X-1 & $ 9.73196\times 10^{14}$  & $ 4.70994\times 10^{14}$  & $ 3.14025\times 10^{13}$ &0.185044  & 0.259969 & 0.0322674   \\

     4U 1538-52 & $ 6.11753\times 10^{14}$  & $ 4.23975\times 10^{14}$  & $ 8.04988\times 10^{12}$ &0.100152  & 0.118246 & 0.0131587   \\

     LMC X-4 & $ 6.60065\times 10^{14}$  & $ 4.28857\times 10^{14}$  & $ 1.03359\times 10^{13}$ &0.116381  &0.141655  &0.0156589    \\
     
     Cen X-3 & $ 8.2047\times 10^{14}$  & $ 4.47979\times 10^{14}$  & $2.00348 \times 10^{13}$ &0.157966  &0.209067  & 0.0244188   \\

     PSR J1903+327 & $ 9.16552\times 10^{14}$  & $ 4.63369\times 10^{14}$  & $ 2.68042\times 10^{13}$ &0.175408  &0.241127  &0.0292446    \\

     4U 1820-30 & $ 8.68725\times 10^{14}$  & $ 4.56407\times 10^{14}$  & $ 2.32442\times 10^{13}$ &0.166766  & 0.224928 & 0.0267567   \\

     EXO 1785-248 & $ 8.45694\times 10^{14}$  & $ 4.71661\times 10^{14}$  & $ 1.96387\times 10^{13}$ &0.15306  &0.200487  &0.023222    \\

\hline
\hline
\end{tabular}
\end{center}
\end{table}

\section{Stability analysis and equilibrium of forces}\label{stable}

\subsection{Casuality Constraint via Herrera's cracking approach}

\begin{figure}[H]
    \centering
        \includegraphics[scale=.6]{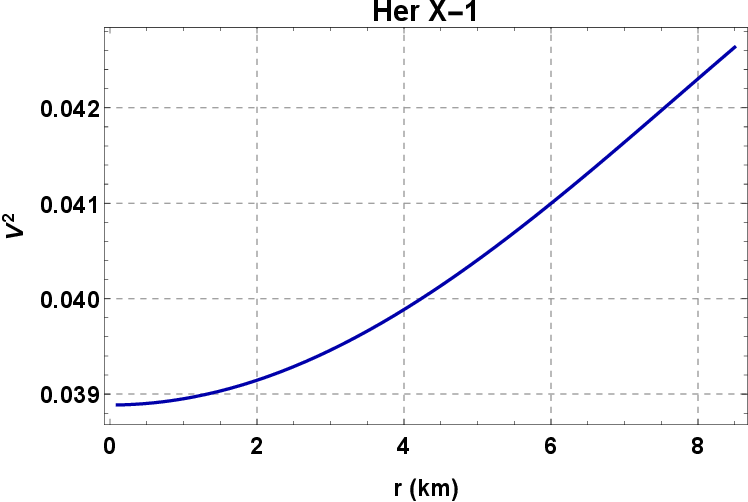}
        \includegraphics[scale=.6]{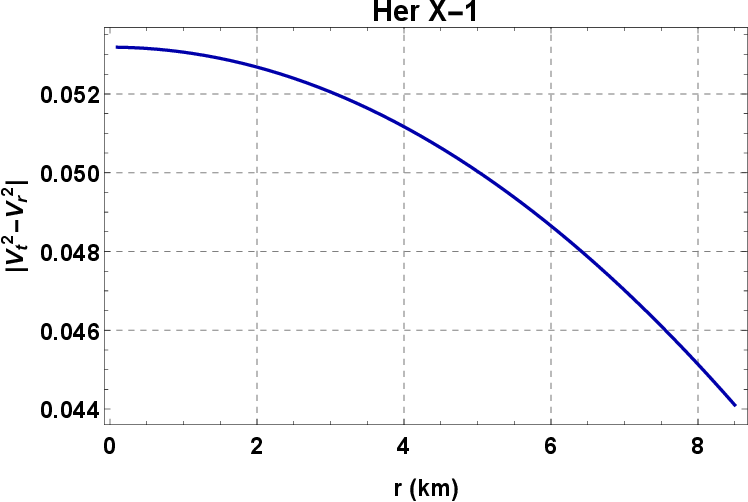}
        \caption{Squared sound velocity and the stability factor $|V_t^2-V_r^2|$ are plotted against radius `r'.}\label{sv}
\end{figure}

In this paragraph, we will look at another crucial "physical acceptability condition" for realistic models: the causality condition, which uses the sound velocity together with Herrera's cracking approach. The causality criterion can be used to evaluate the stability of the model and guarantee its physical feasibility. According to the causality constraint, there must always be a non-negative time-like interval between any two events in spacetime. This matter suggests that information and signals cannot travel faster than light. This means that the square of the sound speed $V^2$ =$\frac{dp}{d\rho}$ must be less than unity for a physically viable model \cite{Herrera:1992lwz, Abreu:2007ew}, which can be validated from the Fig.~(\ref{sv}) for the present model. Hence, the present charged dark energy star model satisfies the causality constraint.
Now, the parameters for effective radial and tangential sound speed components for this present anisotropic model are defined as,
\begin{eqnarray}
V_r^2&=&\frac{dp_r^{\text{eff}}}{d\rho^{\text{eff}}} ~~\rm{and} \label{sp1}\\
V_t^2&=&\frac{dp_t^{\text{eff}}}{d\rho^{\text{eff}}}.\label{sp2}
\end{eqnarray}
Additionally, Herrera proposed the "cracking" (or overturning) method for relativistic star objects under slight radial perturbations \cite{Herrera:1992lwz}. In their study, Abreu et al. employed the concept of cracking \cite{Abreu:2007ew} and introduced the concept of stability factor, which is expressed mathematically as $|V_t^2-V_r^2|<1$. This methodology helps researchers examine the structural integrity of compact systems and identify factors that influence their stability. This criterion is also met throughout this model, as shown by the plot of the stability factor profile in Fig.~(\ref{sv}). Hence, the present model is physically stable and well behaved.

\subsection{Energy conditions}

\begin{figure}[H]
    \centering
        \includegraphics[scale=.65]{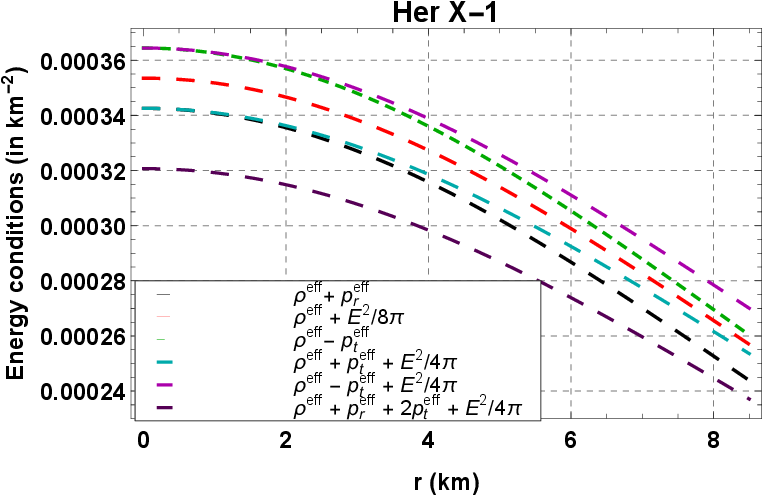}
        \caption{Behavior of all energy conditions inside the stellar interior.}\label{ec}
\end{figure}

Energy conditions (ECs) act as crucial constraints categorized by unique physical properties generated by the stress-energy tensor. Researchers use these limitations to assess the feasibility of different cosmic events. These conditions correspond to gravitational theory by relating the energy density and pressure in spacetime. They also define matter and field properties.\\
Now, we will examine the ECs of a charged fluid sphere model related to anisotropic dark energy using relativistic classical field theories of gravitation. In GR, ECs refer to local inequalities that govern the relationship between matter-energy density and pressure within specific limits. An anisotropic-charged fluid sphere should satisfy the four conventional, model-independent, pointwise ECs known as (i) null energy condition (NEC), (ii) weak energy condition (WEC), (iii) strong energy condition (SEC), and (iv) dominant energy condition (DEC) \cite{bondi1947spherically, witten1981new, visser1997energy, Andreasson:2008xw, garcia2011energy}. For present model, the relevant effective ECs must hold simultaneously across the stellar model, as characterized by the following inequalities:
\begin{itemize}
\item NEC:~$\rho^{\text{eff}}+ p_r^{\text{eff}} \geq 0,~\rho^{\text{eff}}+ p_t^{\text{eff}} + \frac{E^2}{4\pi} \geq 0$, 
\item WEC:~$\rho^{\text{eff}}+ p_r^{\text{eff}} \geq 0,~\rho^{\text{eff}}+ p_t^{\text{eff}} + \frac{E^2}{4\pi} \geq 0,~ \rho^{\text{eff}} + \frac{E^2}{8\pi}  \geq 0$, 
\item SEC:~$\rho^{\text{eff}}+ p_r^{\text{eff}} \geq 0,~\rho^{\text{eff}}+ p_t^{\text{eff}} + \frac{E^2}{4\pi} \geq 0,~\rho^{\text{eff}} + p_r^{\text{eff}} + 2p_t^{\text{eff}}+ \frac{E^2}{4\pi} \geq 0$, 
\item DEC:~$\rho^{\text{eff}}-p_r^{\text{eff}} + \frac{E^2}{4\pi} \geq 0,~ \rho^{\text{eff}}-p_t^{\text{eff}} \geq 0,~\rho^{\text{eff}} + \frac{E^2}{8\pi} \geq 0$.
\end{itemize}
Fig.~(\ref{ec}) shows that the proposed dark energy star model complies with all ECs for all $r$.  Consequently, our model is physically realistic.

\subsection{The Equilibrium of Hydrodynamic Forces via Modified TOV equation}

Now we will look at the hydrostatic equilibrium of the current model in the presence of dark energy. The Tolman-Oppenheimer-Volkofi (TOV) equation \cite{Ghezzi:2009ct, ponce1993limiting} models the equilibrium of a self-gravitating, spherically symmetric compact object with gravitational and internal pressure forces. Using the TOV equation, which is defined as follows, we will verify that this stellar model is in equilibrium when subjected to different forces:
\begin{equation}\label{tov1}
-\frac{M_G(r)(\rho+p)}{r}e^{(\beta-\eta)/2}-\frac{dp}{dr}-\frac{dp_r^{D}}{dr}+\frac{2}{r}(p_t^{D}-p_r^{D})+\sigma Ee^{\beta/2}=0,
\end{equation}
where $M_G(r)$ is effective gravitational mass inside a fluid sphere of radius $r$ as obtained from the Tolman-Whittaker formula \cite{PhysRev.35.875} and Einstein's field equations, is expressed as,
\begin{equation}\label{tov2}
M_G(r)=\frac{1}{2}r^2 \eta' e^{(\eta - \beta)/2}.
\end{equation}
Now, substituting the analytical expression of $M_G(r)$ in Eqn. \eqref{tov1}, we finally obtain,
\begin{equation}\label{tov1a}
-\frac{\eta'}{2}(\rho+p)-\frac{dp}{dr}-\frac{dp_r^{D}}{dr}+\frac{2}{r}(p_t^{D}-p_r^{D})+\sigma Ee^{\beta/2}=0.
\end{equation}
The above modified TOV Eqn. (\ref{tov2}) can also be expressed as,
\begin{equation}
F_g + F_h + F_d + F_e=0,
\end{equation}
where $F_g, F_h$, $F_e$ corresponds to the specific forces as gravitational, hydrostatic-gradient, and electrical force respectively whereas an additional force term $F_d$ arises due to dark energy given as follows:
\begin{eqnarray}
\text{Gravitational force:}~ F_g&=&-\frac{\eta'}{2}(\rho+p) \\
\text{Hydrostatic-gradient force:}~ F_h&=& -\frac{dp}{dr}\\
\text{Dark energy force:}~ F_d&=& -\frac{dp_r^{D}}{dr}+\frac{2}{r}(p_t^{D}-p_r^{D})\\
\text{Electric force:}~ F_e&=& \sigma Ee^{\beta/2}.
\end{eqnarray}
\begin{figure}[H]
    \centering
        \includegraphics[scale=.6]{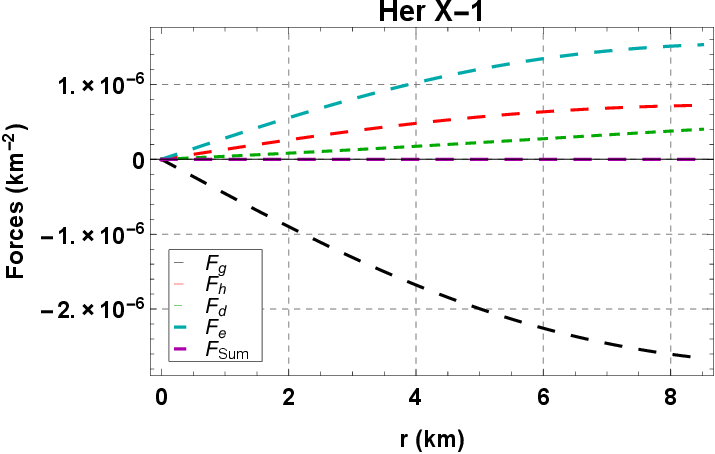}
        \caption{Behavior of different forces, such as the gravitational force $F_g$, hydrostatic gradient force $F_h$, dark energy force $F_d$ and electric force $F_e$ respectively versus radial coordinate $r$. Here $F_{\text{Sum}}=F_g + F_h +F_d + F_e$.}
    \label{tov}
\end{figure}
We can visually verify the behavior of the generalized TOV equations using Fig.~(\ref{tov}), and the equilibrium of the system is met by all forces in the system, i.e., the net force (sum of all forces) equal to zero. This balance of forces ensures the stability of the proposed model. The static equilibrium condition confirms that the forces acting on the system are counterbalanced, preventing gravitational collapse or expansion, thus maintaining a stable configuration for the compact object.\\
In this scenario, it is also observed that the gravitational ($F_g$) is only negative, while the hydrostatic gradient ($F_h$), dark energy forces ($F_d$) and electric forces ($F_e$) are positive. We also plotted the overall influence of all four forces $F_{\text{Sum}}=F_g + F_h +F_d + F_e$ to adequately justify their counterbalance effect and see that it vanishes (magenta line in Fig.~(\ref{tov}). This implies that our current model achieves a static equilibrium.

\subsection{Equation of state parameter: Zel'dovich condition}

\begin{figure}[H]
    \centering
        \includegraphics[scale=.63]{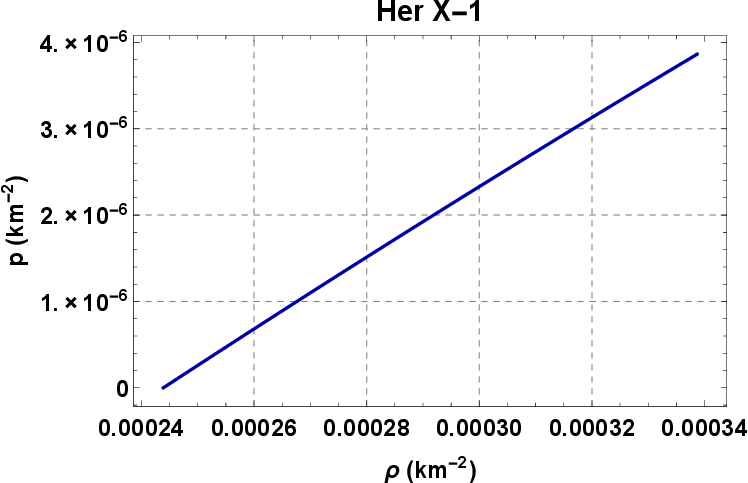}
        \includegraphics[scale=.6]{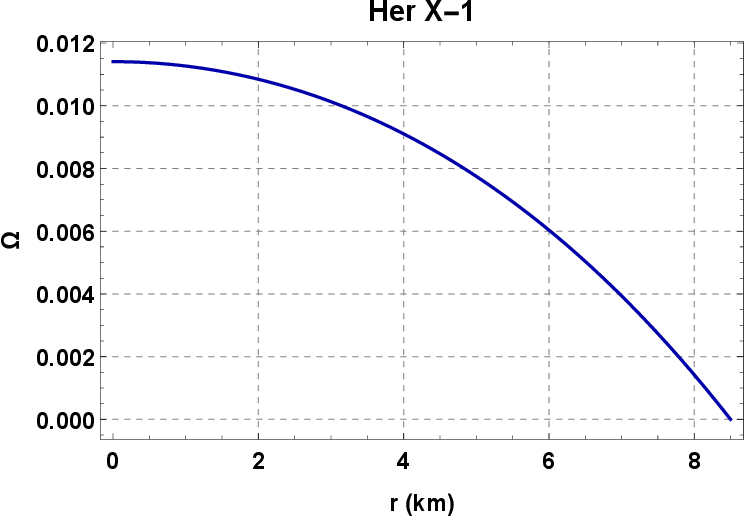}
        \caption{(i)Correlation of pressure $p$ with density $\rho$ and (ii) the variation of $\Omega (r)$ are shown inside the stellar interior.}\label{eos}
\end{figure}

The Zel'dovich criterion for stellar stability \cite{shapiro2008black, zeldovich1971relativistic, zel1972relativistic, l1962equation} states that the pressure-density ratio $\Omega(r)=\frac{p(r)}{\rho(r)}$ should be within $(0,1)$ throughout the stellar interior, mathematically, $0<\Omega<1$. This stability criterion is used mainly in GR and astrophysics to check whether a stellar configuration is physically acceptable. Fig.~(\ref{eos}) shows that this model satisfies the Zel'dovich ratio for all $r < \mathbb{R}$. This ratio illustrates the concept of the equation of state parameter, which is a vital aspect in determining stellar formations. In addition, we can check from Fig.~(\ref{eos}) that the normal baryonic density and pressure demonstrate a linear correlation, which supports the consistency of the model.

\section{Discussion and Concluding Remarks}\label{con}

In this work, we have developed a new model of a charged strange star coupled with anisotropic dark energy by employing the Adler-Finch-Skea solution under Einstein gravity, while strictly satisfying the Karmarkar embedding condition. The model is constructed by effectively combining baryonic matter, electric charge, and a dark energy component modeled with an EoS proportional to the ordinary matter density. These models exhibit stability, static equilibrium, smooth matching of internal and exterior spacetimes, and satisfy energy criteria.

Now, employing the Karmarkar condition allowed us to derive a tractable form of the radial metric potential, with the temporal component chosen based on Adler's methodology. By using observational data from the compact star candidate Her X-1, we determined the constants in the metric through smooth matching with the exterior Reissner-Nordstr\"om spacetime using Darmois-Israel junction conditions. The primary difference between this kind of research and the original dark energy stellar model is that the shell is adjacent to or replaces the event horizon, while the core is de-Sitter.

Several studies have previously been conducted on dark energy stars, a mixture of ordinary fluid matter and dark energy. Shvartsman \cite{1971JETP...33..475S} suggests that stars also can carry electric charges. So we provide a model of a singularity-free charged dark energy star. The charged dark energy star model has become more astrophysically relevant for a variety of reasons, including its potential as a black hole substitute. Additionally, it has been discovered that dark energy and anisotropic stress can assist in stabilizing stars against gravitational collapse by exerting a repulsive force on the dark energy environment that resembles an electric charge.\par
We present a simpler numerical technique for integrating the stellar structure equations from the star's core to its boundary. The compactness, structural characteristics, and M-R relations of the stars were calculated and compared to the well-known theoretical TOV model. The results obtained have been analyzed numerically as well as graphically throughout this paper. Several physical aspects of the star configuration have been investigated to develop a feasible solution for the dark energy star model.

The key findings of this work are summarized as follows:

\begin{itemize}
  \item The graphical representations of $e^{\eta(r)}$ and $e^{-\beta(r)}$ in Fig.~(\ref{metric}) clearly demonstrate that they are finite, non-negative, and singularity-free over the radius of stars, which is the basic condition to construct physically acceptable models of compact stellar objects.

 \item  Graphical representations have been utilized to show the profiles of matter density and pressure components. Fig.~(\ref{rho}) shows that the ordinary matter-energy density $\rho(r)$ and pressure $p(r)$ are continuous and positively finite within the star, with a gradual decline towards the surface. They vanish at the stellar boundary. Both density and pressure are maximum in the core, indicating the occurrence of very compact cores. Electric field intensity ($E^2$) increases with $r$.  It remains positive and monotonically increasing across the fluid sphere.

 \item In Fig.~(\ref{grad}), the gradient components of matter-energy density and pressure show a negative trend, moving from zero to a negative region as $r$ approaches from the center to boundary. The pressure and density gradients vanish at the center ($r=0$) and remain negative throughout the stellar structure. Furthermore, double derivatives assume negative values at the center (in Fig.~(\ref{ddr})), implying that both density and pressure have their highest values at the center of the stellar object.

 \item  This model depicts the repulsive nature of dark energy, resulting in negative dark pressure components and positive energy density $(\rho^{D})$ (See Fig.~\ref{dark}) throughout the configuration.
 
 \item Fig.~(\ref{mass}) demonstrates that the mass function and compactness factor are regular across the stellar region and steadily increase with 'r'.
 
  \item Another crucial feature of compact objects is their gravitational redshift, which indicates how much light is shifted to longer wavelengths by the strong gravitational field. In Fig.~(\ref{red}), we can see that the gravitational (or internal) redshift $(z_g)$ is minimum near the surface and highest in the core. Whereas, the compactness of the compact object, which is determined by its mass-to-radius ratio, determines the surface redshift. However, the surface redshift $(z_s)$ acts exactly opposite to $z_g$.
  
  \item The proposed charged dark energy stellar model also meets the causality requirement, which means that the square of the normal baryonic sound velocity $V^2$ is within $[0,1]$ as well as the stability factor derived from the cracking technique is less than 1 (from Fig.~(\ref{sv})) within the stellar body, resulting in a realistic, physically stable, and well-behaved model throughout the range.
  
  \item Fig.~(\ref{ec}) shows a clear trend of all energy conditions for this solution, which is satisfactory. These conditions are always positive across the entire region of the star and steadily decrease nearer the surface after peaking at the core, confirming the feasibility of the proposed solution. These conditions are very essential to guarantee that the compact object model depicts a stable and physically significant object.

  \item  The TOV equation governs the hydrostatic equilibrium condition, which guarantees that the compact object is balanced under all forces. We observe from Fig.~(\ref{tov}) that when the sum of all acting forces $F_g$, $F_h$, $F_d$, and $F_e$ equals zero, the system attains static equilibrium. This balance is essential for maintaining the structural integrity of the proposed dark energy stellar model.
  
  \item In Fig.~(\ref{eos}), we examine the EoS parameter using the Zel'dovich condition for stability. The EoS for the dark enery star demonstrates a linear correlation for $\rho$ and $p$ in the figure, which supports the consistency of the model even though the star comprises electric charge, normal baryonic matter, and dark energy. These linear characteristics hold true throughout the stellar interior, particularly in the range $0 < \Omega(r) < 1$, reinforcing the realistic nature of this anisotropic charged dark energy star model. They are particularly significant in the central region of the star. It follows that the solution of this suggested model is physically well-behaved and that the matter content is non-exotic in nature. Overall, this work demonstrates that the Adler-Finch-Skea solution coupled with the Karmarkar condition provides a robust and physically consistent framework for describing charged dark energy stars, contributing significantly to our understanding of compact stellar objects influenced by dark energy effects.
  
\end{itemize}

These findings extend our limited exploration of dark energy stars, especially for double-fluid models in Adler-Finch-Skea spacetime. Given all of the significant findings, we can finally conclude that we can construct a physically acceptable, stable, and singularity-free generalized model for charged strange stars using the dark energy EoS, which is suitable for studying dark energy stars within the Einstein gravity framework. Our extensive physical analysis reveals that the dark energy density and negative dark pressures provide a repulsive force that effectively balances gravitational collapse. The model opens avenues for future investigations into the astrophysical implications of such exotic objects and their potential observational signatures.

\section*{Acknowledgements} Pramit Rej is thankful to the Inter-University Centre for Astronomy and Astrophysics (IUCAA), Pune, Government of India, for providing Visiting Associateship. I would also like to express my deepest gratitude to our little cute daughter, Aishiki, whose innocent smiles and boundless joy have been a continual source of inspiration and motivation throughout this research work. This work is dedicated to her, with love and hope for a bright future. 

\section*{Declarations}
\textbf{Funding:} The author did not receive any funding in the form of financial aid or grant from any institution or organization for the present research work.\par
\textbf{Data Availability Statement:} The results are obtained
via purely theoretical calculations and can be verified analytically; thus this manuscript has no associated data, or the data will not be deposited. \par
\textbf{Conflicts of Interest:} The authors have no financial interest or involvement which is relevant by any means to the content of this study.

\bibliography{de_references}

\begin{thebibliography}{70}
\expandafter\ifx\csname natexlab\endcsname\relax\def\natexlab#1{#1}\fi
\expandafter\ifx\csname bibnamefont\endcsname\relax
  \def\bibnamefont#1{#1}\fi
\expandafter\ifx\csname bibfnamefont\endcsname\relax
  \def\bibfnamefont#1{#1}\fi
\expandafter\ifx\csname citenamefont\endcsname\relax
  \def\citenamefont#1{#1}\fi
\expandafter\ifx\csname url\endcsname\relax
  \def\url#1{\texttt{#1}}\fi
\expandafter\ifx\csname urlprefix\endcsname\relax\def\urlprefix{URL }\fi
\providecommand{\bibinfo}[2]{#2}
\providecommand{\eprint}[2][]{\url{#2}}

\bibitem[{\citenamefont{Lobo}(2006)}]{Lobo:2005uf}
\bibinfo{author}{\bibfnamefont{F.~S.~N.} \bibnamefont{Lobo}}, \bibinfo{journal}{Class. Quant. Grav.} \textbf{\bibinfo{volume}{23}}, \bibinfo{pages}{1525} (\bibinfo{year}{2006}), \eprint{gr-qc/0508115}.

\bibitem[{\citenamefont{Perlmutter et~al.}(1999)}]{SupernovaCosmologyProject:1998vns}
\bibinfo{author}{\bibfnamefont{S.}~\bibnamefont{Perlmutter}} \bibnamefont{et~al.} (\bibinfo{collaboration}{Supernova Cosmology Project}), \bibinfo{journal}{Astrophys. J.} \textbf{\bibinfo{volume}{517}}, \bibinfo{pages}{565} (\bibinfo{year}{1999}), \eprint{astro-ph/9812133}.

\bibitem[{\citenamefont{Beltracchi and Gondolo}(2019)}]{beltracchi2019formation}
\bibinfo{author}{\bibfnamefont{P.}~\bibnamefont{Beltracchi}} \bibnamefont{and} \bibinfo{author}{\bibfnamefont{P.}~\bibnamefont{Gondolo}}, \bibinfo{journal}{Physical Review D} \textbf{\bibinfo{volume}{99}}, \bibinfo{pages}{044037} (\bibinfo{year}{2019}).

\bibitem[{\citenamefont{Chapline}(2004)}]{Chapline:2004jfp}
\bibinfo{author}{\bibfnamefont{G.}~\bibnamefont{Chapline}}, \bibinfo{journal}{eConf} \textbf{\bibinfo{volume}{C041213}}, \bibinfo{pages}{0205} (\bibinfo{year}{2004}), \eprint{astro-ph/0503200}.

\bibitem[{\citenamefont{Riess et~al.}(2019)\citenamefont{Riess, Casertano, Yuan, Macri, and Scolnic}}]{Riess:2019cxk}
\bibinfo{author}{\bibfnamefont{A.~G.} \bibnamefont{Riess}}, \bibinfo{author}{\bibfnamefont{S.}~\bibnamefont{Casertano}}, \bibinfo{author}{\bibfnamefont{W.}~\bibnamefont{Yuan}}, \bibinfo{author}{\bibfnamefont{L.~M.} \bibnamefont{Macri}}, \bibnamefont{and} \bibinfo{author}{\bibfnamefont{D.}~\bibnamefont{Scolnic}}, \bibinfo{journal}{Astrophys. J.} \textbf{\bibinfo{volume}{876}}, \bibinfo{pages}{85} (\bibinfo{year}{2019}), \eprint{1903.07603}.

\bibitem[{\citenamefont{Sushkov}(2005)}]{sushkov2005wormholes}
\bibinfo{author}{\bibfnamefont{S.}~\bibnamefont{Sushkov}}, \bibinfo{journal}{Physical Review D} \textbf{\bibinfo{volume}{71}}, \bibinfo{pages}{043520} (\bibinfo{year}{2005}).

\bibitem[{\citenamefont{Bibi et~al.}(2016)\citenamefont{Bibi, Feroze, and Siddiqui}}]{bibi2016solution}
\bibinfo{author}{\bibfnamefont{R.}~\bibnamefont{Bibi}}, \bibinfo{author}{\bibfnamefont{T.}~\bibnamefont{Feroze}}, \bibnamefont{and} \bibinfo{author}{\bibfnamefont{A.~A.} \bibnamefont{Siddiqui}}, \bibinfo{journal}{Canadian Journal of Physics} \textbf{\bibinfo{volume}{94}}, \bibinfo{pages}{758} (\bibinfo{year}{2016}).

\bibitem[{\citenamefont{Feng et~al.}(2008)\citenamefont{Feng, Wang, Abdalla, and Su}}]{feng2008observational}
\bibinfo{author}{\bibfnamefont{C.}~\bibnamefont{Feng}}, \bibinfo{author}{\bibfnamefont{B.}~\bibnamefont{Wang}}, \bibinfo{author}{\bibfnamefont{E.}~\bibnamefont{Abdalla}}, \bibnamefont{and} \bibinfo{author}{\bibfnamefont{R.}~\bibnamefont{Su}}, \bibinfo{journal}{Physics Letters B} \textbf{\bibinfo{volume}{665}}, \bibinfo{pages}{111} (\bibinfo{year}{2008}).

\bibitem[{\citenamefont{Karmarkar}(1948)}]{karmarkar1948gravitational}
\bibinfo{author}{\bibfnamefont{K.}~\bibnamefont{Karmarkar}}, in \emph{\bibinfo{booktitle}{Proceedings of the Indian academy of sciences-section A}} (\bibinfo{organization}{Springer}, \bibinfo{year}{1948}), vol.~\bibinfo{volume}{27}, pp. \bibinfo{pages}{56--60}.

\bibitem[{\citenamefont{Adler}(1974)}]{Adler:1974dn}
\bibinfo{author}{\bibfnamefont{R.~J.} \bibnamefont{Adler}}, \bibinfo{journal}{J. Math. Phys.} \textbf{\bibinfo{volume}{15}}, \bibinfo{pages}{727} (\bibinfo{year}{1974}), \bibinfo{note}{[Erratum: J.Math.Phys. 17, 158 (1976)]}.

\bibitem[{\citenamefont{Finch and Skea}(1989)}]{finch}
\bibinfo{author}{\bibfnamefont{M.~R.} \bibnamefont{Finch}} \bibnamefont{and} \bibinfo{author}{\bibfnamefont{J.~E.~F.} \bibnamefont{Skea}}, \bibinfo{journal}{Class. Quantum Grav.} \textbf{\bibinfo{volume}{6}}, \bibinfo{pages}{467} (\bibinfo{year}{1989}).

\bibitem[{\citenamefont{Ray et~al.}(2003)\citenamefont{Ray, Espindola, Malheiro, Lemos, and Zanchin}}]{ray2003electrically}
\bibinfo{author}{\bibfnamefont{S.}~\bibnamefont{Ray}}, \bibinfo{author}{\bibfnamefont{A.~L.} \bibnamefont{Espindola}}, \bibinfo{author}{\bibfnamefont{M.}~\bibnamefont{Malheiro}}, \bibinfo{author}{\bibfnamefont{J.~P.} \bibnamefont{Lemos}}, \bibnamefont{and} \bibinfo{author}{\bibfnamefont{V.~T.} \bibnamefont{Zanchin}}, \bibinfo{journal}{Physical Review D} \textbf{\bibinfo{volume}{68}}, \bibinfo{pages}{084004} (\bibinfo{year}{2003}).

\bibitem[{\citenamefont{Chan et~al.}(2009{\natexlab{a}})\citenamefont{Chan, da~Silva, and Villas~da Rocha}}]{chan2009star}
\bibinfo{author}{\bibfnamefont{R.}~\bibnamefont{Chan}}, \bibinfo{author}{\bibfnamefont{M.}~\bibnamefont{da~Silva}}, \bibnamefont{and} \bibinfo{author}{\bibfnamefont{J.~F.} \bibnamefont{Villas~da Rocha}}, \bibinfo{journal}{General Relativity and Gravitation} \textbf{\bibinfo{volume}{41}}, \bibinfo{pages}{1835} (\bibinfo{year}{2009}{\natexlab{a}}).

\bibitem[{\citenamefont{Chan et~al.}(2009{\natexlab{b}})\citenamefont{Chan, Da~Silva, and Villas Da~Rocha}}]{chan2009anisotropic}
\bibinfo{author}{\bibfnamefont{R.}~\bibnamefont{Chan}}, \bibinfo{author}{\bibfnamefont{M.}~\bibnamefont{Da~Silva}}, \bibnamefont{and} \bibinfo{author}{\bibfnamefont{J.~F.} \bibnamefont{Villas Da~Rocha}}, \bibinfo{journal}{Modern Physics Letters A} \textbf{\bibinfo{volume}{24}}, \bibinfo{pages}{1137} (\bibinfo{year}{2009}{\natexlab{b}}).

\bibitem[{\citenamefont{Lobo and Arellano}(2007)}]{lobo2007gravastars}
\bibinfo{author}{\bibfnamefont{F.~S.} \bibnamefont{Lobo}} \bibnamefont{and} \bibinfo{author}{\bibfnamefont{A.~V.} \bibnamefont{Arellano}}, \bibinfo{journal}{Classical and Quantum Gravity} \textbf{\bibinfo{volume}{24}}, \bibinfo{pages}{1069} (\bibinfo{year}{2007}).

\bibitem[{\citenamefont{Chan et~al.}(2011)\citenamefont{Chan, Da~Silva, and Rocha}}]{chan2011gravastars}
\bibinfo{author}{\bibfnamefont{R.}~\bibnamefont{Chan}}, \bibinfo{author}{\bibfnamefont{M.}~\bibnamefont{Da~Silva}}, \bibnamefont{and} \bibinfo{author}{\bibfnamefont{P.}~\bibnamefont{Rocha}}, \bibinfo{journal}{General Relativity and Gravitation} \textbf{\bibinfo{volume}{43}}, \bibinfo{pages}{2223} (\bibinfo{year}{2011}).

\bibitem[{\citenamefont{Bertolami and Paramos}(2005)}]{bertolami2005chaplygin}
\bibinfo{author}{\bibfnamefont{O.}~\bibnamefont{Bertolami}} \bibnamefont{and} \bibinfo{author}{\bibfnamefont{J.}~\bibnamefont{Paramos}}, \bibinfo{journal}{Physical Review D} \textbf{\bibinfo{volume}{72}}, \bibinfo{pages}{123512} (\bibinfo{year}{2005}).

\bibitem[{\citenamefont{Cattoen and Faber}(2005)}]{cattoen2005visser}
\bibinfo{author}{\bibfnamefont{C.}~\bibnamefont{Cattoen}} \bibnamefont{and} \bibinfo{author}{\bibfnamefont{T.}~\bibnamefont{Faber}}, \bibinfo{journal}{Class. Quantum Gravity} p. \bibinfo{pages}{4189} (\bibinfo{year}{2005}).

\bibitem[{\citenamefont{Gupta and Maurya}(2011)}]{gupta2011class}
\bibinfo{author}{\bibfnamefont{Y.}~\bibnamefont{Gupta}} \bibnamefont{and} \bibinfo{author}{\bibfnamefont{S.~K.} \bibnamefont{Maurya}}, \bibinfo{journal}{Astrophysics and Space Science} \textbf{\bibinfo{volume}{331}}, \bibinfo{pages}{135} (\bibinfo{year}{2011}).

\bibitem[{\citenamefont{Kiess}(2012)}]{kiess2012exact}
\bibinfo{author}{\bibfnamefont{T.~E.} \bibnamefont{Kiess}}, \bibinfo{journal}{Astrophysics and Space Science} \textbf{\bibinfo{volume}{339}}, \bibinfo{pages}{329} (\bibinfo{year}{2012}).

\bibitem[{\citenamefont{Takisa and Maharaj}(2013)}]{takisa2013some}
\bibinfo{author}{\bibfnamefont{P.~M.} \bibnamefont{Takisa}} \bibnamefont{and} \bibinfo{author}{\bibfnamefont{S.}~\bibnamefont{Maharaj}}, \bibinfo{journal}{General Relativity and Gravitation} \textbf{\bibinfo{volume}{45}}, \bibinfo{pages}{1951} (\bibinfo{year}{2013}).

\bibitem[{\citenamefont{Malaver}(2017)}]{malaver2017new}
\bibinfo{author}{\bibfnamefont{M.}~\bibnamefont{Malaver}}, \bibinfo{journal}{International Journal of Systems Science and Applied Mathematics} \textbf{\bibinfo{volume}{2}}, \bibinfo{pages}{93} (\bibinfo{year}{2017}).

\bibitem[{\citenamefont{Malaver}(2018)}]{malaver2018generalized}
\bibinfo{author}{\bibfnamefont{M.}~\bibnamefont{Malaver}}, \bibinfo{journal}{World Scientific News} \textbf{\bibinfo{volume}{92}}, \bibinfo{pages}{327} (\bibinfo{year}{2018}).

\bibitem[{\citenamefont{Sunzu et~al.}(2014)\citenamefont{Sunzu, Maharaj, and Ray}}]{sunzu2014quark}
\bibinfo{author}{\bibfnamefont{J.~M.} \bibnamefont{Sunzu}}, \bibinfo{author}{\bibfnamefont{S.~D.} \bibnamefont{Maharaj}}, \bibnamefont{and} \bibinfo{author}{\bibfnamefont{S.}~\bibnamefont{Ray}}, \bibinfo{journal}{Astrophysics and Space Science} \textbf{\bibinfo{volume}{354}}, \bibinfo{pages}{517} (\bibinfo{year}{2014}).

\bibitem[{\citenamefont{Nazar et~al.}(2025)\citenamefont{Nazar, Majeed, Abbas, Ashraf, Ibraheem, and Davletov}}]{Nazar2025-gl}
\bibinfo{author}{\bibfnamefont{H.}~\bibnamefont{Nazar}}, \bibinfo{author}{\bibfnamefont{A.}~\bibnamefont{Majeed}}, \bibinfo{author}{\bibfnamefont{G.}~\bibnamefont{Abbas}}, \bibinfo{author}{\bibfnamefont{A.}~\bibnamefont{Ashraf}}, \bibinfo{author}{\bibfnamefont{A.~A.} \bibnamefont{Ibraheem}}, \bibnamefont{and} \bibinfo{author}{\bibfnamefont{E.}~\bibnamefont{Davletov}}, \bibinfo{journal}{Chin. J. Phys.} \textbf{\bibinfo{volume}{96}}, \bibinfo{pages}{1083} (\bibinfo{year}{2025}).

\bibitem[{\citenamefont{Shahzad et~al.}(2025)\citenamefont{Shahzad, Habib, Amjad, Ashraf, Hakami, Boukhris, Alsaiari, and Al-Buriahi}}]{Shahzad:2025fzk}
\bibinfo{author}{\bibfnamefont{M.~R.} \bibnamefont{Shahzad}}, \bibinfo{author}{\bibfnamefont{W.}~\bibnamefont{Habib}}, \bibinfo{author}{\bibfnamefont{M.}~\bibnamefont{Amjad}}, \bibinfo{author}{\bibfnamefont{A.}~\bibnamefont{Ashraf}}, \bibinfo{author}{\bibfnamefont{A.~H.} \bibnamefont{Hakami}}, \bibinfo{author}{\bibfnamefont{I.}~\bibnamefont{Boukhris}}, \bibinfo{author}{\bibfnamefont{N.~S.} \bibnamefont{Alsaiari}}, \bibnamefont{and} \bibinfo{author}{\bibfnamefont{M.~S.} \bibnamefont{Al-Buriahi}}, \bibinfo{journal}{Phys. Dark Univ.} \textbf{\bibinfo{volume}{49}}, \bibinfo{pages}{101964} (\bibinfo{year}{2025}).

\bibitem[{\citenamefont{Ditta et~al.}(2025)\citenamefont{Ditta, Ashraf, Maurya, Ali, Alimova, and Atamurotov}}]{Ditta2025-yc}
\bibinfo{author}{\bibfnamefont{A.}~\bibnamefont{Ditta}}, \bibinfo{author}{\bibfnamefont{A.}~\bibnamefont{Ashraf}}, \bibinfo{author}{\bibfnamefont{S.~K.} \bibnamefont{Maurya}}, \bibinfo{author}{\bibfnamefont{A.}~\bibnamefont{Ali}}, \bibinfo{author}{\bibfnamefont{A.}~\bibnamefont{Alimova}}, \bibnamefont{and} \bibinfo{author}{\bibfnamefont{F.}~\bibnamefont{Atamurotov}}, \bibinfo{journal}{Phys. Dark Universe} \textbf{\bibinfo{volume}{50}}, \bibinfo{pages}{102071} (\bibinfo{year}{2025}).

\bibitem[{\citenamefont{Mustafa et~al.}(2020{\natexlab{a}})\citenamefont{Mustafa, Zubair, Waheed, and Tiecheng}}]{Mustafa:2020yux}
\bibinfo{author}{\bibfnamefont{G.}~\bibnamefont{Mustafa}}, \bibinfo{author}{\bibfnamefont{M.}~\bibnamefont{Zubair}}, \bibinfo{author}{\bibfnamefont{S.}~\bibnamefont{Waheed}}, \bibnamefont{and} \bibinfo{author}{\bibfnamefont{X.}~\bibnamefont{Tiecheng}}, \bibinfo{journal}{Eur. Phys. J. C} \textbf{\bibinfo{volume}{80}}, \bibinfo{pages}{26} (\bibinfo{year}{2020}{\natexlab{a}}).

\bibitem[{\citenamefont{Mustafa et~al.}(2020{\natexlab{b}})\citenamefont{Mustafa, Shamir, and Tie-Cheng}}]{Mustafa:2020jln}
\bibinfo{author}{\bibfnamefont{G.}~\bibnamefont{Mustafa}}, \bibinfo{author}{\bibfnamefont{M.~F.} \bibnamefont{Shamir}}, \bibnamefont{and} \bibinfo{author}{\bibfnamefont{X.}~\bibnamefont{Tie-Cheng}}, \bibinfo{journal}{Phys. Rev. D} \textbf{\bibinfo{volume}{101}}, \bibinfo{pages}{104013} (\bibinfo{year}{2020}{\natexlab{b}}), \eprint{2005.03997}.

\bibitem[{\citenamefont{Ghezzi}(2011)}]{Ghezzi:2009ct}
\bibinfo{author}{\bibfnamefont{C.~R.} \bibnamefont{Ghezzi}}, \bibinfo{journal}{Astrophys. Space Sci.} \textbf{\bibinfo{volume}{333}}, \bibinfo{pages}{437} (\bibinfo{year}{2011}), \eprint{0908.0779}.

\bibitem[{\citenamefont{Pandey and Sharma}(1982)}]{pandey}
\bibinfo{author}{\bibfnamefont{S.}~\bibnamefont{Pandey}} \bibnamefont{and} \bibinfo{author}{\bibfnamefont{S.}~\bibnamefont{Sharma}}, \bibinfo{journal}{General Relativity and Gravitation} \textbf{\bibinfo{volume}{14}}, \bibinfo{pages}{113} (\bibinfo{year}{1982}).

\bibitem[{\citenamefont{Ghezzi}(2005)}]{Ghezzi:2005iy}
\bibinfo{author}{\bibfnamefont{C.~R.} \bibnamefont{Ghezzi}}, \bibinfo{journal}{Phys. Rev. D} \textbf{\bibinfo{volume}{72}}, \bibinfo{pages}{104017} (\bibinfo{year}{2005}), \eprint{gr-qc/0510106}.

\bibitem[{\citenamefont{Barreto et~al.}(2007)\citenamefont{Barreto, Rodriguez, Rosales, and Serrano}}]{Barreto:2006cr}
\bibinfo{author}{\bibfnamefont{W.}~\bibnamefont{Barreto}}, \bibinfo{author}{\bibfnamefont{B.}~\bibnamefont{Rodriguez}}, \bibinfo{author}{\bibfnamefont{L.}~\bibnamefont{Rosales}}, \bibnamefont{and} \bibinfo{author}{\bibfnamefont{O.}~\bibnamefont{Serrano}}, \bibinfo{journal}{Gen. Rel. Grav.} \textbf{\bibinfo{volume}{39}}, \bibinfo{pages}{23} (\bibinfo{year}{2007}), \bibinfo{note}{[Erratum: Gen.Rel.Grav. 39, 537--538 (2007)]}, \eprint{gr-qc/0611089}.

\bibitem[{\citenamefont{Das and Ali}(2015)}]{das2015anisotropic}
\bibinfo{author}{\bibfnamefont{K.}~\bibnamefont{Das}} \bibnamefont{and} \bibinfo{author}{\bibfnamefont{N.}~\bibnamefont{Ali}}, \bibinfo{journal}{Astrophysics and Space Science} \textbf{\bibinfo{volume}{356}}, \bibinfo{pages}{57} (\bibinfo{year}{2015}).

\bibitem[{\citenamefont{Reissner}(1916)}]{reissner1916eigengravitation}
\bibinfo{author}{\bibfnamefont{H.}~\bibnamefont{Reissner}}, \bibinfo{journal}{Annalen der Physik} \textbf{\bibinfo{volume}{355}}, \bibinfo{pages}{106} (\bibinfo{year}{1916}).

\bibitem[{\citenamefont{Nordstr{\"o}m}(1918)}]{nordstrom1918energy}
\bibinfo{author}{\bibfnamefont{G.}~\bibnamefont{Nordstr{\"o}m}}, \bibinfo{journal}{Koninklijke Nederlandse Akademie van Wetenschappen Proceedings Series B Physical Sciences} \textbf{\bibinfo{volume}{20}}, \bibinfo{pages}{1238} (\bibinfo{year}{1918}).

\bibitem[{\citenamefont{Abubekerov et~al.}(2008{\natexlab{a}})\citenamefont{Abubekerov, Antokhina, Cherepashchuk, and Shimanskii}}]{Abubekerov:2008inw}
\bibinfo{author}{\bibfnamefont{M.~K.} \bibnamefont{Abubekerov}}, \bibinfo{author}{\bibfnamefont{E.~A.} \bibnamefont{Antokhina}}, \bibinfo{author}{\bibfnamefont{A.~M.} \bibnamefont{Cherepashchuk}}, \bibnamefont{and} \bibinfo{author}{\bibfnamefont{V.~V.} \bibnamefont{Shimanskii}}, \bibinfo{journal}{Astron. Rep.} \textbf{\bibinfo{volume}{52}}, \bibinfo{pages}{379} (\bibinfo{year}{2008}{\natexlab{a}}), \eprint{1201.5519}.

\bibitem[{\citenamefont{Varela et~al.}(2010)\citenamefont{Varela, Rahaman, Ray, Chakraborty, and Kalam}}]{Varela:2010mf}
\bibinfo{author}{\bibfnamefont{V.}~\bibnamefont{Varela}}, \bibinfo{author}{\bibfnamefont{F.}~\bibnamefont{Rahaman}}, \bibinfo{author}{\bibfnamefont{S.}~\bibnamefont{Ray}}, \bibinfo{author}{\bibfnamefont{K.}~\bibnamefont{Chakraborty}}, \bibnamefont{and} \bibinfo{author}{\bibfnamefont{M.}~\bibnamefont{Kalam}}, \bibinfo{journal}{Phys. Rev. D} \textbf{\bibinfo{volume}{82}}, \bibinfo{pages}{044052} (\bibinfo{year}{2010}), \eprint{1004.2165}.

\bibitem[{\citenamefont{Abubekerov et~al.}(2008{\natexlab{b}})\citenamefont{Abubekerov, Antokhina, Cherepashchuk, and Shimanskii}}]{Abubekerov:2012yj}
\bibinfo{author}{\bibfnamefont{M.~K.} \bibnamefont{Abubekerov}}, \bibinfo{author}{\bibfnamefont{E.~A.} \bibnamefont{Antokhina}}, \bibinfo{author}{\bibfnamefont{A.~M.} \bibnamefont{Cherepashchuk}}, \bibnamefont{and} \bibinfo{author}{\bibfnamefont{V.~V.} \bibnamefont{Shimanskii}}, \bibinfo{journal}{Astron. Rep.} \textbf{\bibinfo{volume}{52}}, \bibinfo{pages}{379} (\bibinfo{year}{2008}{\natexlab{b}}), \eprint{1201.5519}.

\bibitem[{\citenamefont{Rawls et~al.}(2011)\citenamefont{Rawls, Orosz, McClintock, Torres, Bailyn, and Buxton}}]{Rawls:2011jw}
\bibinfo{author}{\bibfnamefont{M.~L.} \bibnamefont{Rawls}}, \bibinfo{author}{\bibfnamefont{J.~A.} \bibnamefont{Orosz}}, \bibinfo{author}{\bibfnamefont{J.~E.} \bibnamefont{McClintock}}, \bibinfo{author}{\bibfnamefont{M.~A.~P.} \bibnamefont{Torres}}, \bibinfo{author}{\bibfnamefont{C.~D.} \bibnamefont{Bailyn}}, \bibnamefont{and} \bibinfo{author}{\bibfnamefont{M.~M.} \bibnamefont{Buxton}}, \bibinfo{journal}{Astrophys. J.} \textbf{\bibinfo{volume}{730}}, \bibinfo{pages}{25} (\bibinfo{year}{2011}), \eprint{1101.2465}.

\bibitem[{\citenamefont{Freire et~al.}(2011)}]{Freire:2010tf}
\bibinfo{author}{\bibfnamefont{P.~C.~C.} \bibnamefont{Freire}} \bibnamefont{et~al.}, \bibinfo{journal}{Mon. Not. Roy. Astron. Soc.} \textbf{\bibinfo{volume}{412}}, \bibinfo{pages}{2763} (\bibinfo{year}{2011}), \eprint{1011.5809}.

\bibitem[{\citenamefont{Guver et~al.}(2010)\citenamefont{Guver, Ozel, Cabrera-Lavers, and Wroblewski}}]{Guver:2008gc}
\bibinfo{author}{\bibfnamefont{T.}~\bibnamefont{Guver}}, \bibinfo{author}{\bibfnamefont{F.}~\bibnamefont{Ozel}}, \bibinfo{author}{\bibfnamefont{A.}~\bibnamefont{Cabrera-Lavers}}, \bibnamefont{and} \bibinfo{author}{\bibfnamefont{P.}~\bibnamefont{Wroblewski}}, \bibinfo{journal}{Astrophys. J.} \textbf{\bibinfo{volume}{712}}, \bibinfo{pages}{964} (\bibinfo{year}{2010}), \eprint{0811.3979}.

\bibitem[{\citenamefont{Ozel et~al.}(2009)\citenamefont{Ozel, Guver, and Psaltis}}]{Ozel:2008kb}
\bibinfo{author}{\bibfnamefont{F.}~\bibnamefont{Ozel}}, \bibinfo{author}{\bibfnamefont{T.}~\bibnamefont{Guver}}, \bibnamefont{and} \bibinfo{author}{\bibfnamefont{D.}~\bibnamefont{Psaltis}}, \bibinfo{journal}{Astrophys. J.} \textbf{\bibinfo{volume}{693}}, \bibinfo{pages}{1775} (\bibinfo{year}{2009}), \eprint{0810.1521}.

\bibitem[{\citenamefont{Delgaty and Lake}(1998)}]{Delgaty:1998uy}
\bibinfo{author}{\bibfnamefont{M.~S.~R.} \bibnamefont{Delgaty}} \bibnamefont{and} \bibinfo{author}{\bibfnamefont{K.}~\bibnamefont{Lake}}, \bibinfo{journal}{Comput. Phys. Commun.} \textbf{\bibinfo{volume}{115}}, \bibinfo{pages}{395} (\bibinfo{year}{1998}), \eprint{gr-qc/9809013}.

\bibitem[{\citenamefont{Pant}(2010)}]{Pant:2010iub}
\bibinfo{author}{\bibfnamefont{N.}~\bibnamefont{Pant}}, \bibinfo{journal}{Astrophys. Space Sci.} \textbf{\bibinfo{volume}{331}}, \bibinfo{pages}{633} (\bibinfo{year}{2010}).

\bibitem[{\citenamefont{Chu and Tan}(2022)}]{Chu:2021uec}
\bibinfo{author}{\bibfnamefont{C.-S.} \bibnamefont{Chu}} \bibnamefont{and} \bibinfo{author}{\bibfnamefont{H.~S.} \bibnamefont{Tan}}, \bibinfo{journal}{Universe} \textbf{\bibinfo{volume}{8}}, \bibinfo{pages}{250} (\bibinfo{year}{2022}), \eprint{2103.06314}.

\bibitem[{\citenamefont{Darmois}(1927)}]{darmois1927equations}
\bibinfo{author}{\bibfnamefont{G.}~\bibnamefont{Darmois}}, \bibinfo{journal}{Paris France}  (\bibinfo{year}{1927}).

\bibitem[{\citenamefont{Israel}(1966)}]{Israel:1966rt}
\bibinfo{author}{\bibfnamefont{W.}~\bibnamefont{Israel}}, \bibinfo{journal}{Nuovo Cim. B} \textbf{\bibinfo{volume}{44S10}}, \bibinfo{pages}{1} (\bibinfo{year}{1966}), \bibinfo{note}{[Erratum: Nuovo Cim.B 48, 463 (1967)]}.

\bibitem[{\citenamefont{Chandrasekhar}(1984)}]{chandrasekhar1984stars}
\bibinfo{author}{\bibfnamefont{S.}~\bibnamefont{Chandrasekhar}}, \bibinfo{journal}{Science} \textbf{\bibinfo{volume}{226}}, \bibinfo{pages}{497} (\bibinfo{year}{1984}).

\bibitem[{\citenamefont{Peebles and Ratra}(2003)}]{Peebles:2002gy}
\bibinfo{author}{\bibfnamefont{P.~J.~E.} \bibnamefont{Peebles}} \bibnamefont{and} \bibinfo{author}{\bibfnamefont{B.}~\bibnamefont{Ratra}}, \bibinfo{journal}{Rev. Mod. Phys.} \textbf{\bibinfo{volume}{75}}, \bibinfo{pages}{559} (\bibinfo{year}{2003}), \eprint{astro-ph/0207347}.

\bibitem[{\citenamefont{Baum and Frampton}(2007)}]{Baum:2006ee}
\bibinfo{author}{\bibfnamefont{L.}~\bibnamefont{Baum}} \bibnamefont{and} \bibinfo{author}{\bibfnamefont{P.~H.} \bibnamefont{Frampton}}, \bibinfo{journal}{Phys. Rev. Lett.} \textbf{\bibinfo{volume}{98}}, \bibinfo{pages}{071301} (\bibinfo{year}{2007}), \eprint{hep-th/0610213}.

\bibitem[{\citenamefont{Buchdahl}(1959{\natexlab{a}})}]{buchdahl1959general}
\bibinfo{author}{\bibfnamefont{H.~A.} \bibnamefont{Buchdahl}}, \bibinfo{journal}{Physical Review} \textbf{\bibinfo{volume}{116}}, \bibinfo{pages}{1027} (\bibinfo{year}{1959}{\natexlab{a}}).

\bibitem[{\citenamefont{Glendenning}(2012)}]{glendenning2012compact}
\bibinfo{author}{\bibfnamefont{N.~K.} \bibnamefont{Glendenning}}, \emph{\bibinfo{title}{Compact stars: Nuclear physics, particle physics and general relativity}} (\bibinfo{publisher}{Springer Science \& Business Media}, \bibinfo{year}{2012}).

\bibitem[{\citenamefont{Florides}(1983)}]{florides1983complete}
\bibinfo{author}{\bibfnamefont{P.~S.} \bibnamefont{Florides}}, \bibinfo{journal}{Journal of Physics A: Mathematical and General} \textbf{\bibinfo{volume}{16}}, \bibinfo{pages}{1419} (\bibinfo{year}{1983}).

\bibitem[{\citenamefont{Kumar and Bharti}(2022)}]{kumar2022isotropic}
\bibinfo{author}{\bibfnamefont{J.}~\bibnamefont{Kumar}} \bibnamefont{and} \bibinfo{author}{\bibfnamefont{P.}~\bibnamefont{Bharti}}, \bibinfo{journal}{The European Physical Journal Plus} \textbf{\bibinfo{volume}{137}}, \bibinfo{pages}{330} (\bibinfo{year}{2022}).

\bibitem[{\citenamefont{Buchdahl}(1959{\natexlab{b}})}]{Buchdahl:1959zz}
\bibinfo{author}{\bibfnamefont{H.~A.} \bibnamefont{Buchdahl}}, \bibinfo{journal}{Phys. Rev.} \textbf{\bibinfo{volume}{116}}, \bibinfo{pages}{1027} (\bibinfo{year}{1959}{\natexlab{b}}).

\bibitem[{\citenamefont{Herrera}(1992)}]{Herrera:1992lwz}
\bibinfo{author}{\bibfnamefont{L.}~\bibnamefont{Herrera}}, \bibinfo{journal}{Phys. Lett. A} \textbf{\bibinfo{volume}{165}}, \bibinfo{pages}{206} (\bibinfo{year}{1992}).

\bibitem[{\citenamefont{Abreu et~al.}(2007)\citenamefont{Abreu, Hernandez, and Nunez}}]{Abreu:2007ew}
\bibinfo{author}{\bibfnamefont{H.}~\bibnamefont{Abreu}}, \bibinfo{author}{\bibfnamefont{H.}~\bibnamefont{Hernandez}}, \bibnamefont{and} \bibinfo{author}{\bibfnamefont{L.~A.} \bibnamefont{Nunez}}, \bibinfo{journal}{Class. Quant. Grav.} \textbf{\bibinfo{volume}{24}}, \bibinfo{pages}{4631} (\bibinfo{year}{2007}), \eprint{0706.3452}.

\bibitem[{\citenamefont{Bondi}(1947)}]{bondi1947spherically}
\bibinfo{author}{\bibfnamefont{H.}~\bibnamefont{Bondi}}, \bibinfo{journal}{Monthly Notices of the Royal Astronomical Society} \textbf{\bibinfo{volume}{107}}, \bibinfo{pages}{410} (\bibinfo{year}{1947}).

\bibitem[{\citenamefont{Witten}(1981)}]{witten1981new}
\bibinfo{author}{\bibfnamefont{E.}~\bibnamefont{Witten}}, \bibinfo{journal}{Communications in Mathematical Physics} \textbf{\bibinfo{volume}{80}}, \bibinfo{pages}{381} (\bibinfo{year}{1981}).

\bibitem[{\citenamefont{Visser}(1997)}]{visser1997energy}
\bibinfo{author}{\bibfnamefont{M.}~\bibnamefont{Visser}}, \bibinfo{journal}{Science} \textbf{\bibinfo{volume}{276}}, \bibinfo{pages}{88} (\bibinfo{year}{1997}).

\bibitem[{\citenamefont{Andreasson}(2009)}]{Andreasson:2008xw}
\bibinfo{author}{\bibfnamefont{H.}~\bibnamefont{Andreasson}}, \bibinfo{journal}{Commun. Math. Phys.} \textbf{\bibinfo{volume}{288}}, \bibinfo{pages}{715} (\bibinfo{year}{2009}), \eprint{0804.1882}.

\bibitem[{\citenamefont{Garcia et~al.}(2011)\citenamefont{Garcia, Harko, Lobo, and Mimoso}}]{garcia2011energy}
\bibinfo{author}{\bibfnamefont{N.~M.} \bibnamefont{Garcia}}, \bibinfo{author}{\bibfnamefont{T.}~\bibnamefont{Harko}}, \bibinfo{author}{\bibfnamefont{F.~S.} \bibnamefont{Lobo}}, \bibnamefont{and} \bibinfo{author}{\bibfnamefont{J.~P.} \bibnamefont{Mimoso}}, \bibinfo{journal}{Physical Review D} \textbf{\bibinfo{volume}{83}}, \bibinfo{pages}{104032} (\bibinfo{year}{2011}).

\bibitem[{\citenamefont{Ponce~de Leon}(1993)}]{ponce1993limiting}
\bibinfo{author}{\bibfnamefont{J.}~\bibnamefont{Ponce~de Leon}}, \bibinfo{journal}{General relativity and gravitation} \textbf{\bibinfo{volume}{25}}, \bibinfo{pages}{1123} (\bibinfo{year}{1993}).

\bibitem[{\citenamefont{Tolman}(1930)}]{PhysRev.35.875}
\bibinfo{author}{\bibfnamefont{R.~C.} \bibnamefont{Tolman}}, \bibinfo{journal}{Phys. Rev.} \textbf{\bibinfo{volume}{35}}, \bibinfo{pages}{875} (\bibinfo{year}{1930}), \urlprefix\url{https://link.aps.org/doi/10.1103/PhysRev.35.875}.

\bibitem[{\citenamefont{Shapiro and Teukolsky}(2008)}]{shapiro2008black}
\bibinfo{author}{\bibfnamefont{S.~L.} \bibnamefont{Shapiro}} \bibnamefont{and} \bibinfo{author}{\bibfnamefont{S.~A.} \bibnamefont{Teukolsky}}, \emph{\bibinfo{title}{Black holes, white dwarfs, and neutron stars: The physics of compact objects}} (\bibinfo{publisher}{John Wiley \& Sons}, \bibinfo{year}{2008}).

\bibitem[{\citenamefont{Zeldovich and Novikov}(1971)}]{zeldovich1971relativistic}
\bibinfo{author}{\bibfnamefont{Y.~B.} \bibnamefont{Zeldovich}} \bibnamefont{and} \bibinfo{author}{\bibfnamefont{I.~D.} \bibnamefont{Novikov}}, \bibinfo{journal}{Chicago: University of Chicago Press}  (\bibinfo{year}{1971}).

\bibitem[{\citenamefont{Zel'dovich et~al.}(1972)\citenamefont{Zel'dovich, Novikov, and Silk}}]{zel1972relativistic}
\bibinfo{author}{\bibfnamefont{Y.~B.} \bibnamefont{Zel'dovich}}, \bibinfo{author}{\bibfnamefont{I.~D.} \bibnamefont{Novikov}}, \bibnamefont{and} \bibinfo{author}{\bibfnamefont{J.}~\bibnamefont{Silk}}, \bibinfo{journal}{Physics Today} \textbf{\bibinfo{volume}{25}}, \bibinfo{pages}{63} (\bibinfo{year}{1972}).

\bibitem[{\citenamefont{L'DOVICH}(1962)}]{l1962equation}
\bibinfo{author}{\bibfnamefont{Y.}~\bibnamefont{L'DOVICH}}, \bibinfo{journal}{Sov. Phys. JETP} \textbf{\bibinfo{volume}{14}}, \bibinfo{pages}{1609} (\bibinfo{year}{1962}).

\bibitem[{\citenamefont{{Shvartsman}}(1971)}]{1971JETP...33..475S}
\bibinfo{author}{\bibfnamefont{V.~F.} \bibnamefont{{Shvartsman}}}, \bibinfo{journal}{Soviet Journal of Experimental and Theoretical Physics} \textbf{\bibinfo{volume}{33}}, \bibinfo{pages}{475} (\bibinfo{year}{1971}).

\end{thebibliography}

\end{document}